\documentclass[%
 reprint,
amsmath,amssymb,
aps,
prb,
]{revtex4-2}

\usepackage{graphicx}
\usepackage{dcolumn}
\usepackage{bm}
\usepackage{physics}
\usepackage{xcolor}
\usepackage{hyperref}
\usepackage{microtype}
\usepackage{soul}

\usepackage{hyperref}
\usepackage{xcolor}
\hypersetup{
    colorlinks,
    linkcolor={red!30!black},
    citecolor={blue!80!black},
    urlcolor={blue!80!black}
}

\usepackage[caption=false]{subfig}
\usepackage{verbatim}
\usepackage[all,cmtip]{xy}
\usepackage{soul}

\begin{document}

\preprint{APS/123-QED}

\title{Plaquette Models, Cellular Automata, and Measurement-induced Criticality}

\author{Hanchen Liu}
\email{hanchen.liu@bc.edu}
\author{Xiao Chen}
\email{chenaad@bc.edu}

\affiliation{Department of Physics, Boston College, Chestnut Hill, Massachusetts 02467, USA}%

\date{\today}

\begin{abstract}

We present a class of two-dimensional randomized plaquette models, where the multi-spin interaction term, referred to as the plaquette term, is replaced by a single-site spin term with a probability of $1-p$. By varying $p$, we observe a ground state phase transition, or equivalently, a phase transition of the symmetry operator. We find that as we vary $p$, the symmetry operator changes from being extensive to being localized in space. These models can be equivalently understood as 1+1D randomized cellular automaton dynamics, allowing the 2D transition to be interpreted as a 1+1D dynamical absorbing phase transition. In this paper, our primary focus is on the plaquette term with three or five-body interactions, where we explore the universality classes of the transitions. Specifically, for the model with five-body interaction, we demonstrate that it belongs to the same universality class as the measurement-induced entanglement phase transition observed in 1+1D Clifford dynamics, as well as the boundary entanglement transition of the 2D cluster state induced by random bulk Pauli measurements. This work establishes a connection between transitions in classical spin models, cellular automata, and hybrid random circuits.

\end{abstract}

\maketitle

\tableofcontents

\section{Introduction}

Classical Ising spin models with multi-spin interaction have been extensively studied in physics~\cite{mccoy1973two, AdamLipowski_1997, PhysRevE.60.5068, HNishimori_1979, barrat1997p, RevModPhys.51.659}. These models represent a fundamental tool for understanding non-trivial physical phenomena such as phase transitions, critical behaviors, and glassy dynamics. The symmetries associated with spin-flip operations play a crucial role in elucidating the physics of these spin models.

The shape of the spin-flip symmetry varies among different Ising-type classical spin models. In the standard Ising model with two-body interactions, there exists a global spin-flip symmetry. This symmetry arises because the energy remains invariant when all Ising spins in the model are flipped simultaneously~\cite{mccoy1973two}. On the other end, models like the Ising gauge theory~\cite{RevModPhys.51.659}, where the classical spins are located on the edges of the lattice, have local spin-flip symmetries associated with each site.  In this model, flipping spins connected to a specific site does not alter the Hamiltonian. This form of symmetry is also known as local gauge symmetry.

In between the ``global" and ``local" lies  the subsystem symmetry and fractal-shaped symmetry. The subsystem symmetry operates on low-dimensional subsystems within the entire system. This symmetry is evident in models such as the 4-body plaquette Ising models~\cite{PhysRevLett.28.507, mueller2017exact, you2018subsystem}, where it is possible to flip a line or a plane of spins without altering the form of the Hamiltonian. This notion of subsystem symmetry can be further generalized into fractal-shaped symmetry~\cite{devakul2019fractal}, where the spin-flips form a fractal-shaped pattern. One famous example is the Sierpinski triangular shape spin-flip symmetry in the Newman-Moore model~\cite{PhysRevE.60.5068}. 

Motivated by previous works, this work introduces a class of two dimensional $q$-body Ising models. We replace these multi-spin interactions randomly with single-site terms. Specifically, we focus on the cases of $q=3$ and $q=5$, and observe that in these random models, varying the probability $1-p$ of single-site terms induces a phase transition in the ground state. We investigate this transition by analyzing the ground state degeneracy and the associated spin flip symmetry operators, which transform between different ground states. We find that this transition can be characterized by a structural change in the symmetry operator. When $p>p^c$, the symmetry operator becomes non-local and can spread over the entire system. In contrast, when $p<p^c$, the symmetry operator is localized within a finite region. 

Furthermore, we demonstrate that these 2D random spin models can be effectively treated as 1+1D randomized cellular automaton dynamics. Through this approach, we show that the localization transition of the symmetry operator can now be mapped to the absorbing phase transition. When $p>p^c$, a symmetry operator starting from one boundary at time $t=1$ can reach the other boundary even after a long time evolution. Conversely, when $p<p^c$, this boundary symmetry operator quickly vanishes into the bulk.  We find that when $q=3$, this cellular automaton dynamics corresponds to a special limit of the Domany-Kinzel cellular automaton(DKCA)~\cite{PhysRevLett.53.311, kinzel1985phase}, while the dynamics with $q=5$ is a second-order automaton~\cite{MARGOLUS198481} with some random constraints. Interestingly, this $q=5$ model has a notable connection with the measurement-induced entanglement phase transition (MIPT).

The MIPT has been extensively studied in various hybrid random circuits~\cite{PhysRevA.62.062311, li2019measurement, PhysRevB.99.224307, PhysRevLett.125.030505, PhysRevX.10.041020, PhysRevB.101.104302, PhysRevB.103.104306, skinner2019measurement, PRXQuantum.2.030313, PhysRevB.106.144311, bao2022finite,Li_2018}. These studies reveal a generic phase transition from a highly entangled volume-law phase to a disentangled area-law phase as the measurement strength or rate is varied. The universality class of these transitions depends on the specific model. Due to the randomness in these circuits, understanding these criticalities poses a challenging problem. Numerically, the 1+1D hybrid random Clifford circuit has been well studied due to its efficient simulation on classical computers.
 
Besides the $1+1$D hybrid circuits, investigations on the boundary phase transition on holographic tensor networks~\cite{PhysRevB.100.134203}, the random stabilizer tensor networks~\cite{PhysRevB.105.104306}, as well as those boundary criticalities induced by bulk measurements on 2D quantum states~\cite{PhysRevB.106.144311, negari2023measurement, liu2023quantum}, offer a fresh perspective on comprehending the entanglement phase transitions. 

With the aforementioned context in mind, in this paper, we rigorously establishes the equivalence between the phase transition of the symmetry operator in the 2D classical random spin model with $q=5$ and the boundary entanglement phase transition induced by bulk Pauli measurements on 2D cluster states. This boundary transition can be effectively mapped to a 1+1D hybrid Clifford circuit. Numerical investigations confirm that the transitions in all three models belong to the same universality class. Notably, the dynamics of the symmetry operator offers a novel perspective for understanding the phase transition in the 1+1D hybrid Clifford circuit. 

The subsequent sections of this paper are structured as follows. 
In Section~\ref{sec: prelim}, we present the fundamental concepts of random plaquette models, including symmetries and  quantities designed to elucidate the structure of the symmetry group, such as symmetry entropy and mutual information. 
Moving on to Section~\ref{sec: sym}, we examine the ground state phase transitions in random plaquette models without boundary, characterizing these transitions in terms of the structure of the symmetry group.  
In Section~\ref{sec: auto}, we discuss the boundary phase transition of the random plaquette models defined on the cylinder, showing that it can be treated as the dynamical phase transition in randomized cellular automata. This section further establishes the equivalence between Pauli measurement on a stabilizer state and the plaquette models. Moreover, we argue that the phase transition of the random Ising model with $q=5$ falls in the same universality class as the boundary phase transition induced by bulk Pauli measurement on the 2D cluster state and the 1+1D measurement-induced phase transition in random Clifford circuits.
Section~\ref{sec: conclusion} devotes to the conclusion and discussion.

\section{Preliminaries}\label{sec: prelim}

We examine a class of $q$-spin Ising models defined on two-dimensional lattices:
\begin{equation}\label{eq: pm}
H[\sigma] = -\sum_{m_q} J_{m_q} \prod_{i \in m_q} \sigma_i,
\end{equation}
where $\sigma_i\in\{-1,1\}$ represents an Ising spin operator at $i$-th site and $\prod_{i \in m_q} \sigma_i$ describes a group of $q$ neighboring spins that are linked together, referred to as the $m_q$-th plaquette term. The Hamiltonian sums over all these plaquette terms. We set the coupling constant uniformly with $J_{m_q}=1$.  It is worth noting that we could also consider a random coupling strength with $J_{m_q}=\pm 1$, and the ground state physics explored in this paper would remain the same. However, for simplicity, we focus on $J_{m_q}=1$ in this work. One well-known instance is the two-dimensional Ising model, where each plaquette term is a two-body interaction, and its ground state is twofold degenerate. This model also displays a finite-temperature phase transition.

\begin{figure}[ht]
    \centering
    \subfloat[RTPM : $q = \Delta$ and $q = \bullet$]{\label{fig: rand_coup_RTPM}\includegraphics[width=0.2\textwidth]{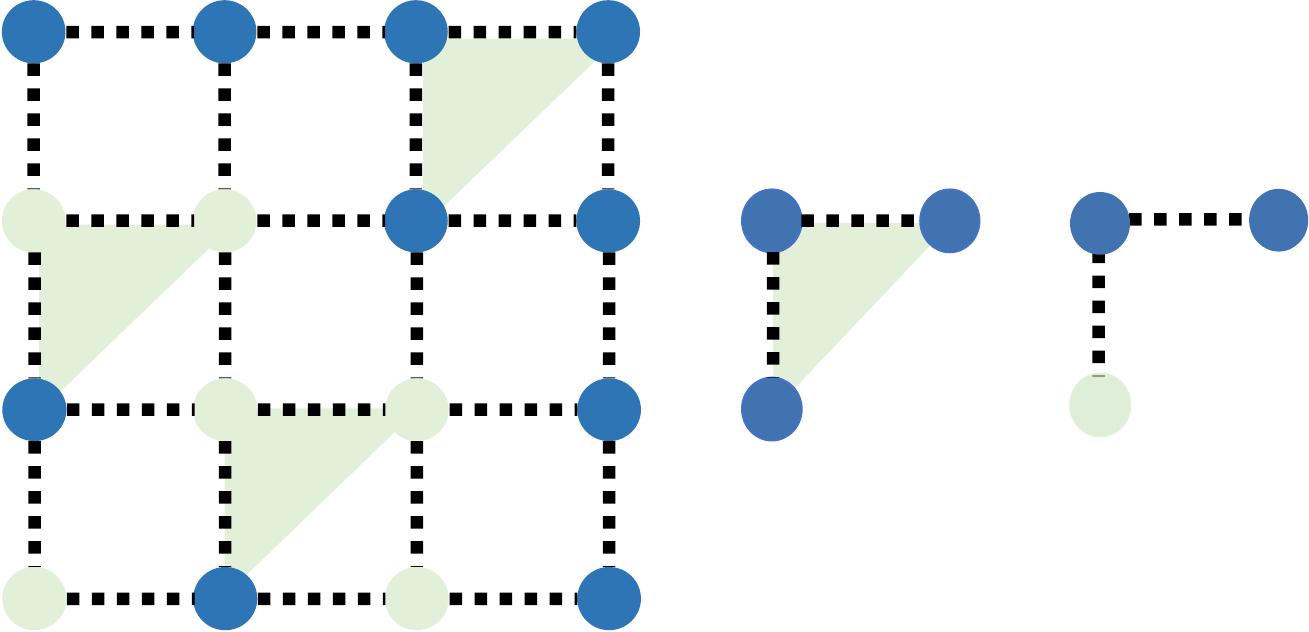}}\qquad
    \subfloat[RXPM: $q = +$ and $q = \bullet$]{\label{fig: rand_coup_RXPM}\includegraphics[width=0.23\textwidth]{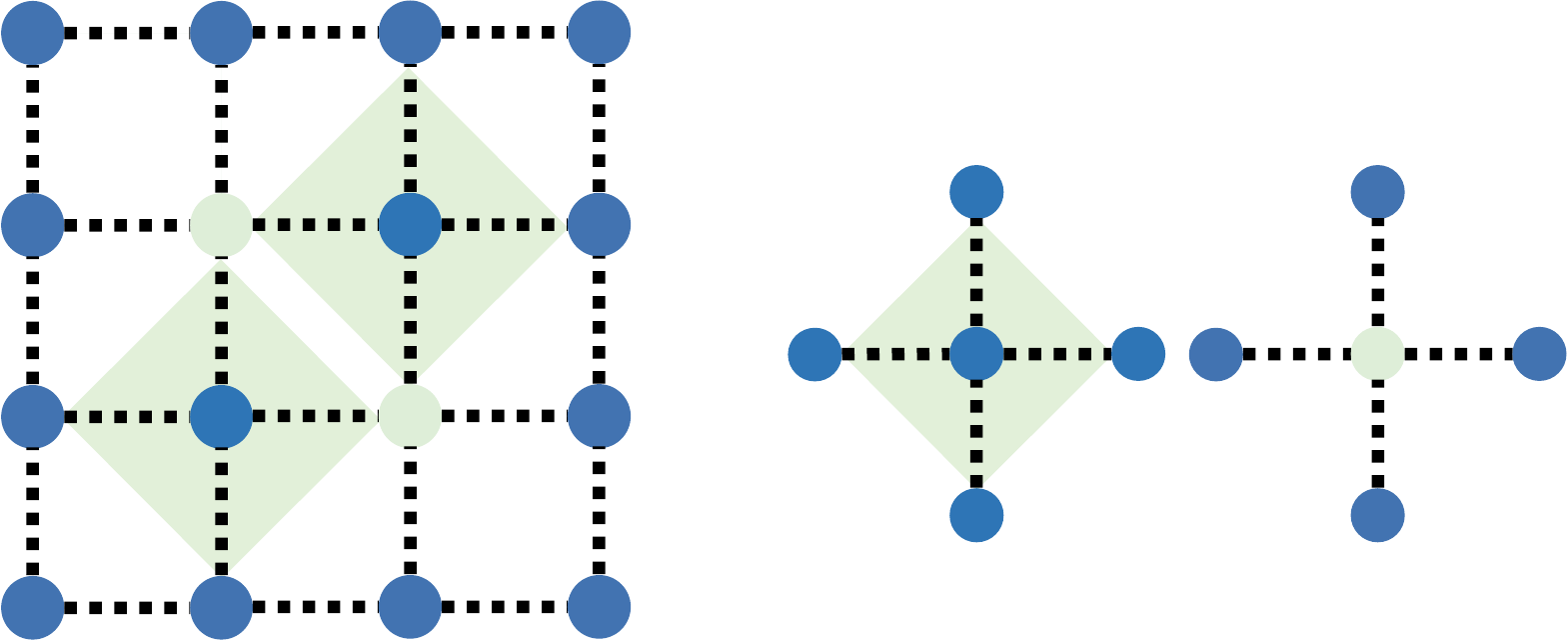}}\qquad
    \caption{\label{fig: RPM} Interaction pattern of the RPM models. }
\end{figure}

In this work, we focus on two spin models with plaquette terms representing three-body and five-body interactions, respectively, defined on a square lattice. At each site, we associate a three/five-body interaction, as shown in Fig.~\ref{fig: rand_coup_RTPM}. We then randomly replace some plaquette terms with single spin terms, resulting in two random plaquette models (RPMs).

The first model is the random triangular plaquette model (RTPM), where each plaquette term is drawn from a binary distribution:
\begin{equation}\label{eq: RTPM}
    \mu(\prod_{i\in m_q} \sigma_i) = \left\{
    \begin{matrix}
        p_3 & q = \Delta\\
        1 - p_3 & q = \bullet\\
    \end{matrix}\right.
\end{equation}
where $p_3$ is the probability of the three-body term denoted as `$q = \Delta$,' and $1 - p_3$ is the probability of taking the single site  $\sigma$, denoted as `$q = \bullet$,' as illustrated in Fig.~\ref{fig: rand_coup_RTPM}. This model converges to the Newman-Moore model in the limit $p_3\to 1$.

Another model we investigate is the random X plaquette model (RXPM), the random couplings of which are taken from 
\begin{equation}\label{eq: RXPM}
    \mu(\prod_{i\in m_q} \sigma_i) = \left\{
    \begin{matrix}
        p_5 & q = + \\
        1 - p_5 & q = \bullet\\
    \end{matrix}\right.
\end{equation}
where $p_5$ is the probability of the five-body coupling, denoted as `$q = +$,' and $1 - p_5$ is again the probability of taking the single site $\sigma$, denoted as `$q = \bullet$,' as illustrated in Fig.~\ref{fig: rand_coup_RXPM}.

\subsection{ Ground state physics}\label{sec: gen_sym}

 \begin{figure*}
    \centering
    \includegraphics[width=0.9\textwidth]{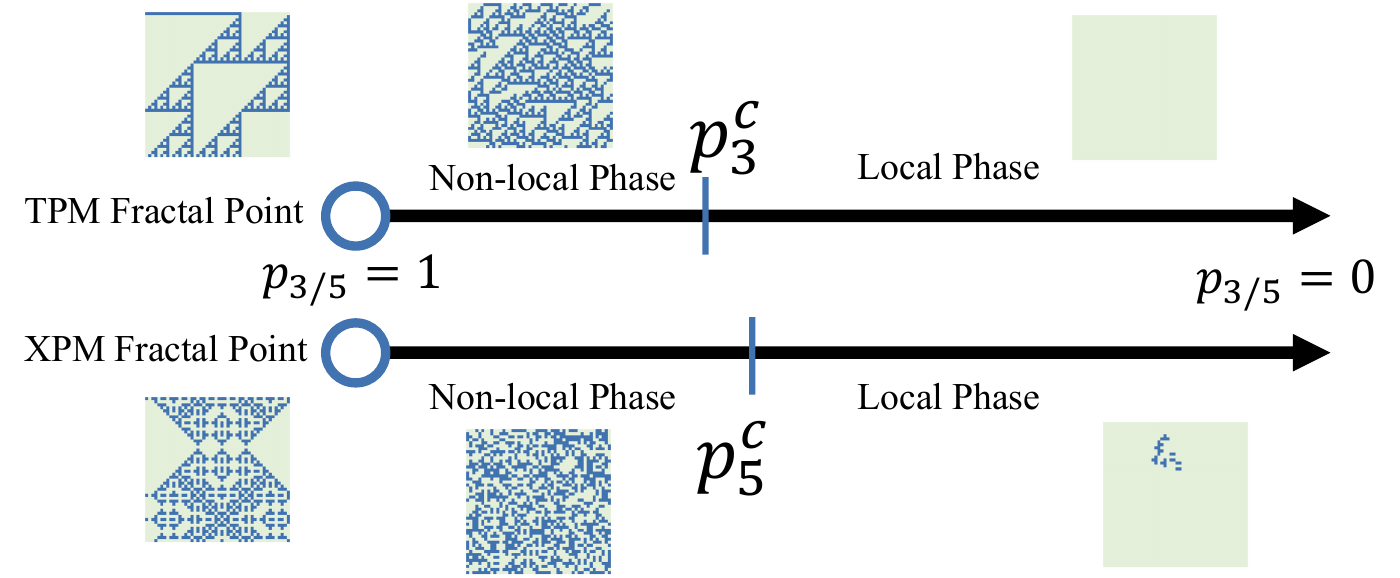}
    \caption{ The phase diagram and snapshots of the symmetry operator of RTPM and RXPM. Blue-colored sites host the spin-flip while the green background represents zeros. The snapshots are taken at $p_{3/5} = 1, 0.9, 0.5$, and the spin-flip symmetry operator $g$ is randomly draw from 
    the full symmetry group $G$.}
    \label{fig: phase_diag}
\end{figure*}

The ground states of these models must satisfy all of the $\prod_{i \in m_q} \sigma_i=1$ defined in the Hamiltonian. There is one trivial ground state where all spins $\sigma=1$. To find other non-trivial ground states, we reformulate this problem in terms of a parity check problem. Through the mapping $\sigma=(-1)^x$ (with $x\in \{0,1\}$), each term is mapped to a parity check
\begin{equation}\label{eq:const}
    \sum_i P_{m_q i} x^i = 0\ \mbox{mod}\ 2
\end{equation}
where $P_{m_q i} = 1$ if site $i$ is in the $m_q$-th plauqette, otherwise $P_{m_q i} = 0$. The index $m_q/i$ runs from $1$ to $Q/N$, with $Q$ being the number of plaquettes and $N$ being the number of sites.

This problem can then be compactly rewritten as 
\begin{align}
    PT=0\ \mbox{mod}\ 2,
\end{align}
where $P$ is a $Q\times N$ parity check matrix and in RTPM/RXPM models, the number of ones per row can either be 1 or 3/5. The solution space is spanned by column vectors of matrix $T$ that is $N\times M$. The matrix is written as
\begin{equation}\label{eq: t_mat}
    T = (\vec{x}_1, \vec{x}_2,\ldots, \vec{x}_M).
\end{equation}
where each column $\vec{x}_{i = 1\ldots M}$ in the $T$ matrix represents an independent solution of the linear system in Eq.~\ref{eq:const}, and $M$ is the total number of independent solutions. Each ground state 
$\vec{x}_{\vec{\alpha}} = \sum_{i=1}^M \alpha_i\vec{x}_i$ is labelled by a binary vector $\vec{\alpha}=(\alpha_1,\cdots,\alpha_M)$ and there are $2^M$ such states.

The $T$ matrix can be directly solved by the standard Gaussian elimination in the following manner. By row reduction, properly relabeling the column indices and removing the rows that are all zeros, we have 
\begin{equation}
    P = (I_r \mid C),
\end{equation}
where $I_r$ denotes the $r\times r$ identity matrix and $C$ is a $r\times (N - r)$ binary matrix, with $r = \rank P$. It is then obvious that 
\begin{equation}
    T= \left(
    \begin{matrix}
        C \\
        I_{M}
    \end{matrix}\right) ,
\end{equation}
as 
\begin{equation}
    (I_r ~ C)\left(
    \begin{matrix}
        C \\
        I_{M}
    \end{matrix}\right) = C + C = 0~\mathrm{mod}~2,
\end{equation}
where $M = N - r$,  and here and in the remaining parts we use $\rank$ to refer to the binary rank.
This algorithm can be realized in polynomial time. Similar problems arise in classical error-correcting codes~\cite{pless1998introduction} and the XORSAT problem~\cite{mezard2009information}. In classical error-correcting codes, the $T$ matrix describes the logic space of classical information, while in XORSAT, the solution space $T$ exhibits interesting phase transitions.

\subsection{Spin-flip symmetries}

We define the spin-flip symmetry $g_{\vec{x}} \in G$ as a tensor product of local spin-flips $X_i$
\begin{equation}
    g_{\vec{x}} = \prod_{i=1}^N X_i^{x_i} \quad\text{with}~ x_i = 0, 1
\end{equation}
that does not change the energy given by Eq.~\ref{eq: pm}:
\begin{equation}
    H[\sigma] = H[g\sigma],
\end{equation}
where the spin-flip $X_i$ flips the classical spin on site $i$ as follows:
\begin{equation}
    X_i \sigma_i = - \sigma_i.
\end{equation}
Here again we use index-$i$ to label the lattice sites.

If $\vec{x}_m$ is a solution of Eq.~\ref{eq:const}, then $g \vec{x}_m$ is also a solution, since the spin-flip symmetry $g$ does not change the energy given by Eq.~\ref{eq: pm} by definition. Thus, we can associate each $\vec{x}_m$ with a $g_m \in G$ by the spin-flips that connect the trivial ground state $\vec{x}_0$, where $\sigma_i = 1$ for all $i$, and a non-trivial solution $\vec{x}_m$ of Eq.~\ref{eq:const}:
\begin{equation}
    \vec{x}_m = g_{\vec{x}_m} \vec{x}_0, \quad g_{\vec{x}_m} \in G.
\end{equation}
The spin-flip symmetry operator can now be written in a more specific form as
\begin{equation}\label{eq: symmtries}
    g_{\vec{x}_m} = \prod_{i=1}^{N} X_i^{x_m^i}
\end{equation}
where $x_m^i$ is the $i$-th entry of vector $\vec x_m$.

The ground state properties are now encoded in the structure of the spin-flip symmetry group. The total number of group elements is now
\begin{equation}
    |G| = 2^M 
\end{equation}
as there are a total of $2^M$ ground states, and the number of independent generators is obtained via

\begin{equation}
    \text{\# of gen.} = \log |G| = M.
\end{equation}
Here and in the remaining parts of this work, we take $\log$ to be 2-based. Another important aspect of $G$ is that it is abelian, which indicates that for any subgroup $G_A \subset G$, we are able to find a group quotient $G/G_A$, the number of elements of which satisfies
\begin{equation}\label{eq: quo}
\begin{aligned}
      \log |G/G_{A}| &= \log (|G| / |G_A|) \\
                    &= \log |G| - \log |G_A|.\\
\end{aligned}
\end{equation}
 and for any subregion $\mathcal{D}$
\begin{equation}\label{eq: rank_nullity}
    \log |G| - \log |G_{\mathcal{D}}| = \rank T_{\overline{\mathcal D}}
\end{equation}
where $T_{\overline{\mathcal D}}$ is the submatrix of $T$ taking only entries associated with domain $\overline{\mathcal{D}}$.
These characteristics of group $G$ will be extensively used in the following sections.

\subsection{Characteristics of symmetry operators} 
In this section, we introduce several quantities to further characterize the ground state spin configurations, or equivalently the structure of the symmetry operators of the RPMs.

Firstly, we define the configuration entropy as the log of the dimension of the full solution space $\vec{x}$ of Eq.~\ref{eq:const}, which is equivalent to the rank of the matrix $T$:
\begin{equation}
    S_{\text{cf}} \equiv \rank T=\log |G|.
\end{equation}
This quantity also refers to the number of independent generators of the symmetry group $G$.

To capture the non-locality of the symmetry operator, we introduce the symmetry entropy with respect to a subregion $A$:
\begin{equation}
\begin{aligned}
    \mathrm{sym}S_A
    &\equiv \log |G / (G_A \cdot G_{\overline{A}})| \\
    &= \log |G| - \log |G_A| - \log |G_{\overline{A}}| \\
    &= \rank T_A + \rank T_{\overline{A}} - \rank T,
\end{aligned}
\end{equation}
where $G_{A/\overline{A}}$ is a subgroup of $G$ containing only elements located inside domain $A$ or its complement $\overline{A}$, and $T_{A/\overline{A}}$ is submatrix of $T$ containing entries associated only with sites in region $A/\overline{A}$. The symmetry entropy $\mathrm{sym}S_A$ thus counts the number of symmetry operators shared by both $A$ and $\overline{A}$. 

Similar to $\mathrm{sym}S_A$, we can define another quantity to characterize the spatial structure of the symmetry operator. This quantity is defined as follows:
\begin{equation}
\begin{aligned}\label{eq: sym_mi_conn}
    \mathrm{sym}I_{AB|C} &\equiv \log \left| G^C/(G_{A}^C \cdot G_B^C) \right| \\
    &= \log |G_C| - \log |G_{AC}| - \log|G_{BC}| + \log |G|  \\
    &= \rank T_A + \rank T_B - \rank T_{AB}
\end{aligned}
\end{equation}
where we divide the system into three subsystems A, B and C with $G^C = G / G_C$, $G_{A}^C = G_{AC} / G_C$, and $G_{B}^C = G_{BC} / G_{C}$. Here, in the second line, we use Eq.~\ref{eq: quo}, and in the third line, we use Eq.~\ref{eq: rank_nullity} together with the fact that $\rank T = \log |G|$. This quantity $\mathrm{sym}I_{AB|C}$ computes the number of symmetry operators that have non-trivial support on both subsystems $A$ and $B$.

\section{Phase transitions in RPMs}\label{sec: sym}

    \begin{figure}[ht]
        \centering
        \subfloat[RTPM operator size]{\label{fig: op_size_TPM}\includegraphics[width=0.24\textwidth]{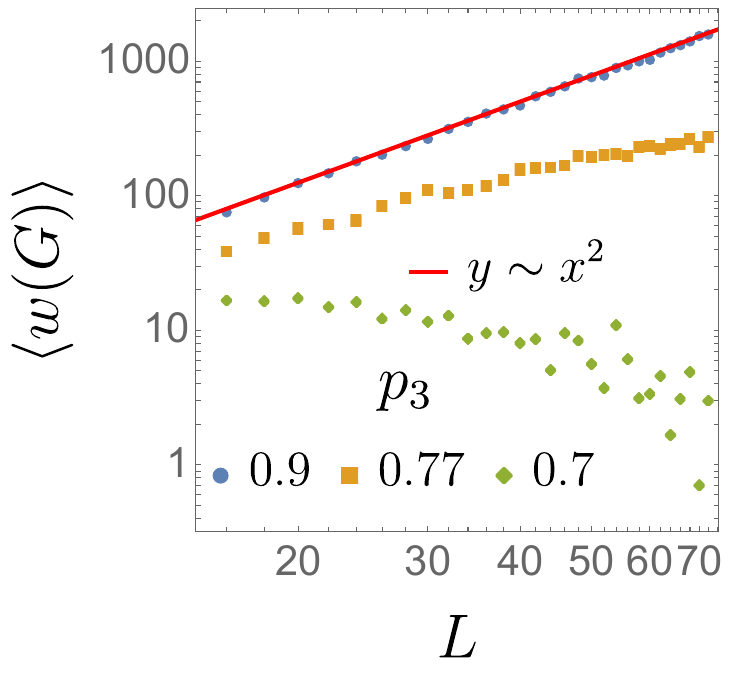}}
        \subfloat[RXPM operator size]{\label{fig: op_size_XPM}\includegraphics[width=0.24\textwidth]{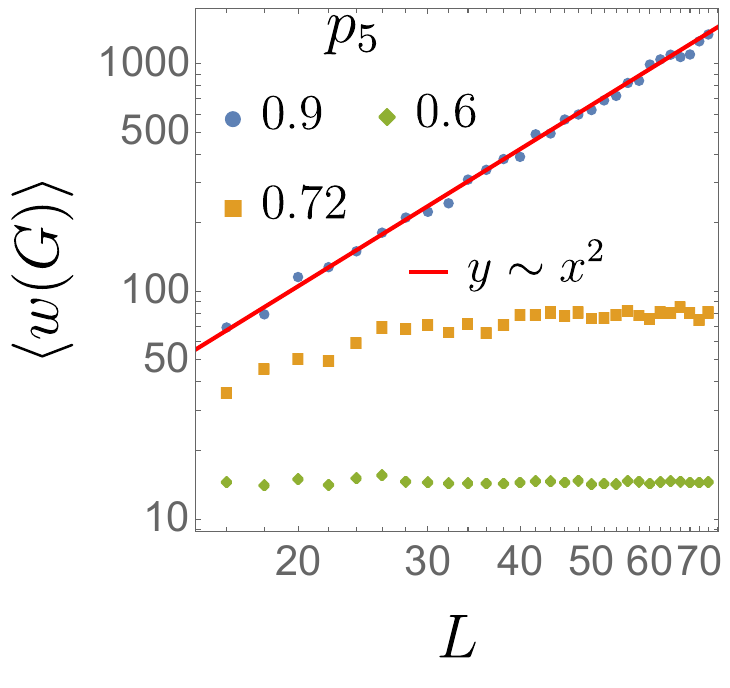}}\qquad
        \caption{\label{fig: op_size} Typical operator size of the spin-flip symmetry operators on a $L \times L$ torus. The operator size $w(G)$ of group $G$ is defined by the Hamming weight of the representation vector $\vec{x}_m$ of a randomly chosen $g_m \in G$, and $\langle\ldots\rangle$ denotes the average of the interaction patterns of RPM models.
        }
    \end{figure}

     We place RTPM and RXPM models on the torus by implementing periodic boundary conditions. We observe that both models exhibit phase transitions as the probabilities $p_3/p_5$ are varied, which are characterized by the structure of the symmetry operator. When $p_3=p_5=1$, both models possess fractal-shaped symmetry operators, with one example shown in Fig.~\ref{fig: phase_diag}. As we depart from this fractal point, the symmetry operator becomes more extensive, occupying $O(L^2)$ number of sites, as present in Fig.~\ref{fig: op_size}. Further reduction of $p_3/p_5$ results in another phase transition. Specifically, for the RTPM model, when $p_3<p_3^c$, the symmetry operator disappears. In contrast, for the RXPM model, symmetry operators persist but only occupy a finite number of sites. We proceed to analyze these phase transitions in both models quantitatively using the quantities introduced in the previous section.

    \begin{figure}[ht]
        \centering
        \subfloat[RTPM $S_{\text{cf}}$]{\label{fig: RTPM_c_entropy}\includegraphics[width=0.22\textwidth]{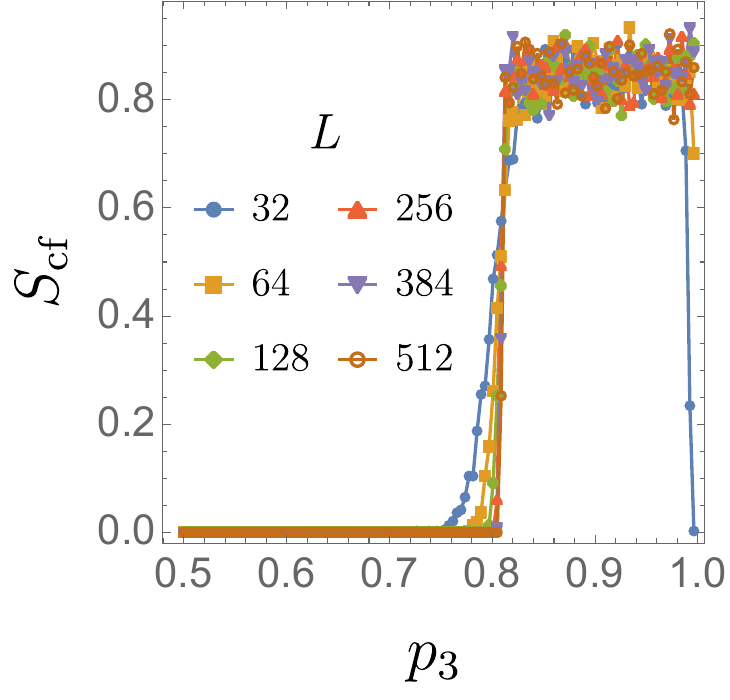}}\qquad
        \subfloat[RTPM $S_{\text{cf}}$ data collapse]{\label{fig: RTPM_c_entropy_dc}\includegraphics[width=0.205\textwidth]{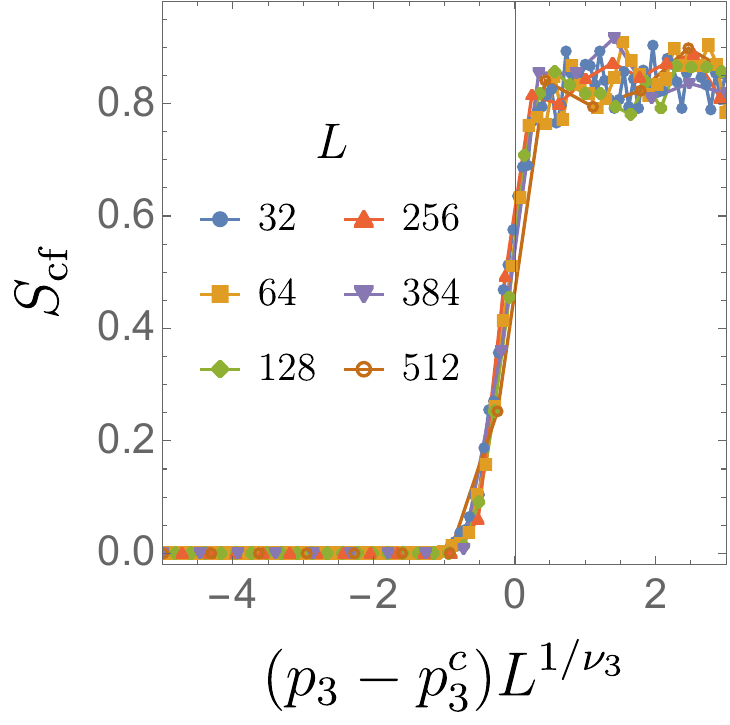}}\qquad
        \subfloat[RXPM $S_{\text{cf}}$]{\label{fig: TPM_c_entropy}\includegraphics[width=0.2\textwidth]{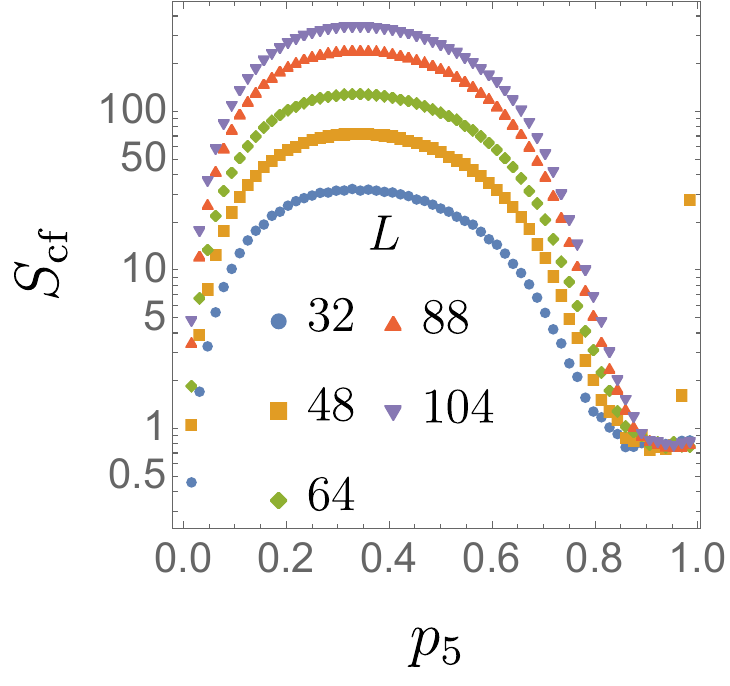}}\qquad
        \subfloat[RXPM $S_{\text{cf}}$ data collapse]{\label{fig: XPM_c_entropy}\includegraphics[width=0.23\textwidth]{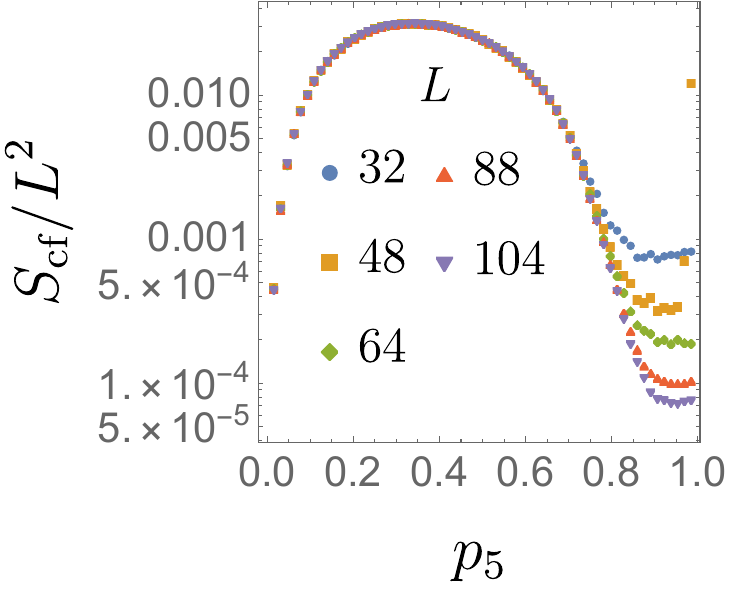}}
        \caption{\label{fig: c_entropy} The data analysis for configuration entropy $ S_{\text{cf}}$ of RTPM defined on a $L \times L^z$ torus with $z = 1.69$ and RXPM defined on a $L\times L$ torus.}
    \end{figure}

    \subsection{Random triangular plaquette model (RTPM)}

    We perform numerical computations to determine the configuration entropy of RTPM as a function of $p_3$. This quantity represents the logarithm of the ground state degeneracy or the number of symmetry operators. We find that when $p_3$ is large, the configuration entropy $S_{\text{cf}}$ remains an $O(1)$ constant and  decreases to 0 as $p_3$ decreases. The critical point $p_3 = 0.81$ is marked by crossing point of $S_{\text{cf}}$ calculated in different $L$ present in Fig.~\ref{fig: RTPM_c_entropy}.

\begin{figure}[ht]
        \centering
        \subfloat[Antipodal line-shaped domains $A,B$, with $L_A = L_B = 1$ for RTPM defined on a $L\times L^z$ torus with $z = 1.697$, where boundaries with the same arrow are glued together]{\label{fig: mi_set_p}\includegraphics[width=0.23\textwidth]{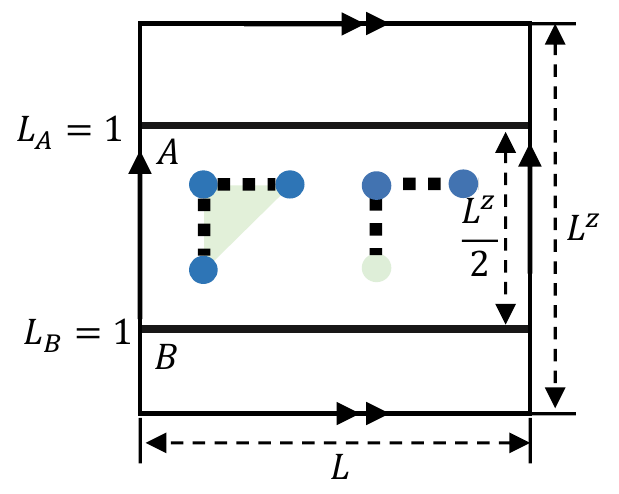}}\qquad
        \subfloat[$\mathrm{sym}I_{AB|C}$ data collapse of RTPM, with $p_3^c = 0.81 $, $\nu_3 = 1.21$.]{\label{fig: RTPM_mi_p}\includegraphics[width=0.18\textwidth]{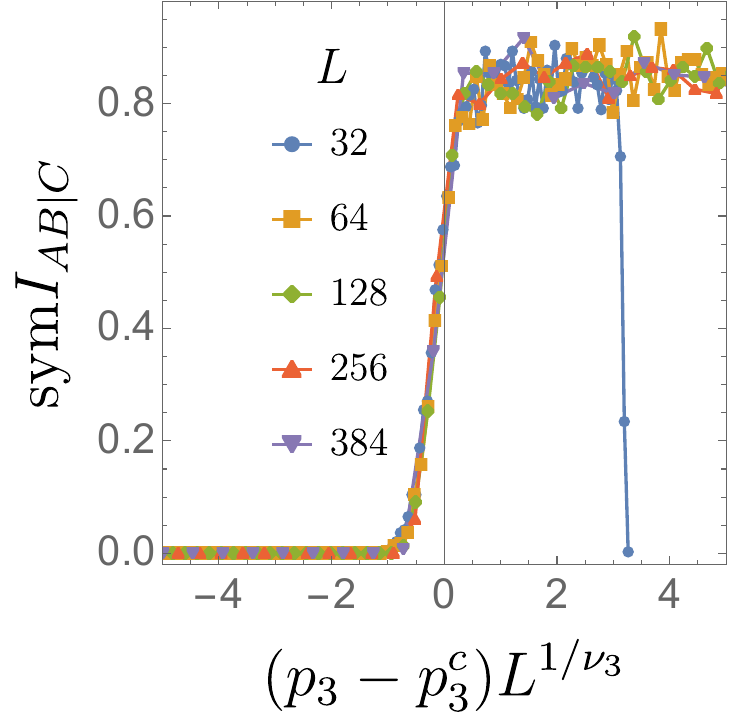}}\qquad
        \subfloat[Antipodal line-shaped domains $A,B$, with $L_A = L_B = 2$ for RXPM defined on a $L\times L$ torus, where boundaries with the same arrow are glued together]{\label{fig: mi_set_X}\includegraphics[width=0.25\textwidth]{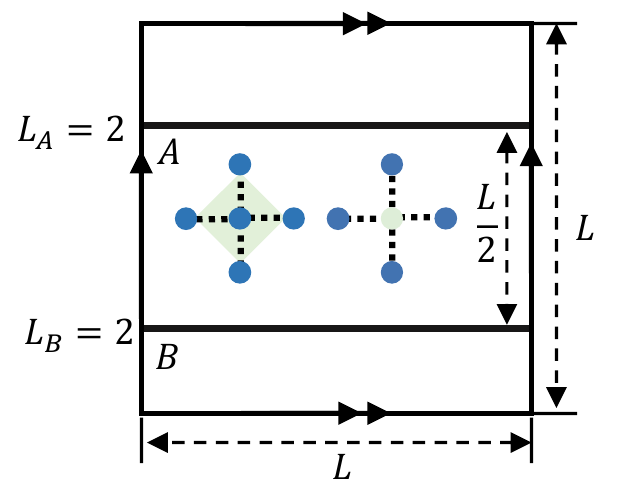}}\qquad
        \subfloat[$\mathrm{sym}I_{AB|C}$ data collapse of RXPM, with $p_5^c = 0.743 $, $\nu_5 = 1.3$.]{\label{fig: RXPM_mi}\includegraphics[width=0.19\textwidth]{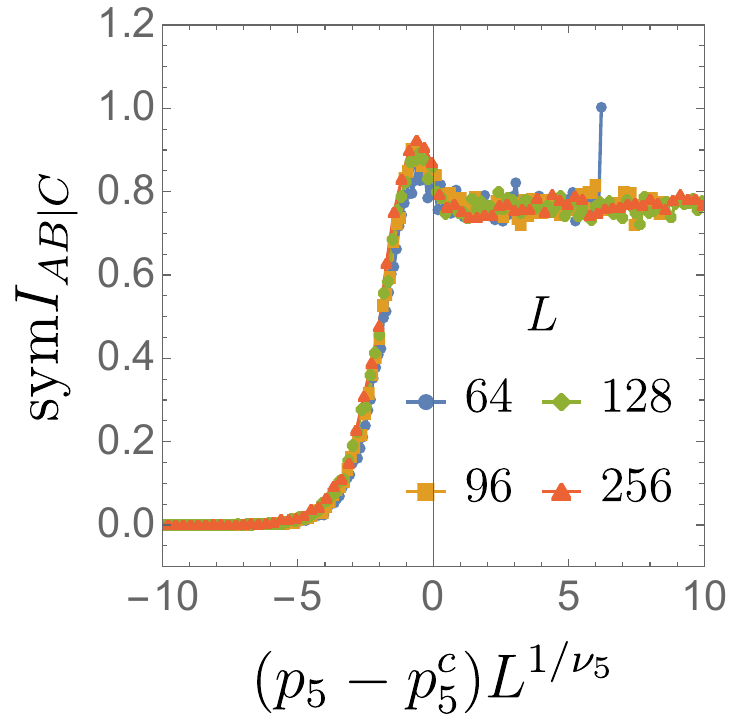}}
        \caption{\label{fig: RPM_mi}  $\mathrm{sym}I_{AB|C}$ of strip domain $A,B$ and $C = \overline{AB}$.}
    \end{figure}

    As shown in Fig.~\ref{fig: RTPM_c_entropy_dc}, around the critical point, we observe that $S_{\text{cf}}$ for different system sizes collapse onto a single function according to the following scaling form:
    \begin{equation}\label{eq: s_cf_dc}
        S_{\text{cf}} \sim f(L^{1/\nu_3}(p - p_c)) 
    \end{equation}
    where $p_3^c \sim 0.81$ and $\nu_3 \sim 1.21$. Here we take the torus to be of size $L \times L^z$ with the geometry shown in Fig.~\ref{fig: mi_set_p}. The parameter $z = 1.697$ is used to capture the anisotropy of RTPM in these two directions. The physical meaning of $z$ and the numerical method to calculate this quantity will be further discussed in Sec.~\ref{sec: auto}.
 The successful collapse of the data implies that in the limit $L\to\infty$,
        \begin{equation}\label{eq: conf_s}
        S_{\text{cf}}\sim \left\{
        \begin{aligned}
            \text{const.} &\quad p_3 > p_3^c\\
            0 &\quad p_3 < p_3^c
        \end{aligned}\right. .
    \end{equation}
   The ground state has finite degeneracy when $p_3>p_3^c$ and has no degeneracy when $p_3<p_3^c$.

    Moreover, we explore the spatial structure of the symmetry operators by computing $\mathrm{sym}I_{AB|C}$. This quantity is defined in Eq.~\ref{eq: sym_mi_conn} and counts the number of symmetry operators which have non-trivial support on both $A$ and $B$. Specifically, we take the geometry as shown in Fig.~\ref{fig: mi_set_p}, where $A$ and $B$ are two antipodal line-shaped domains. When $p_3 > p_3^c$, all generators of the spin-flip symmetry are non-local, and there are $O(1)$ of them,  resulting in $\mathrm{sym}I_{AB|C} = S_{\text{cf}}\sim O(1)$. Conversely, for $p_3 < p_3^c$, there is no ground state degeneracy, and hence no symmetry operator is shared by $A$ and $B$, leading to $\mathrm{sym}I_{AB|C} = 0$.
    
    Through the above analysis, we observe that $S_{\text{cf}}$ and $\mathrm{sym}I_{AB|C}$ are identical in both phases. This is confirmed by the result in Fig.~\ref{fig: RTPM_mi_p}, where we take the same data collapse for $\mathrm{sym}I_{AB|C}$. The non-local structure of the symmetry operator is also confirmed by the independence of $\mathrm{sym}S_A$ as a function of the subsystem size, as demonstrated in Fig.~\ref{fig: RTPM_ee}.

    \subsection{Random X plaquette model (RXPM)}  
\begin{figure}[ht]
        \centering
        \subfloat[Shape of subregion $A$ of RXPM on $L\times L$ torus]{\label{fig: ee_set_X}\includegraphics[width=0.22\textwidth]{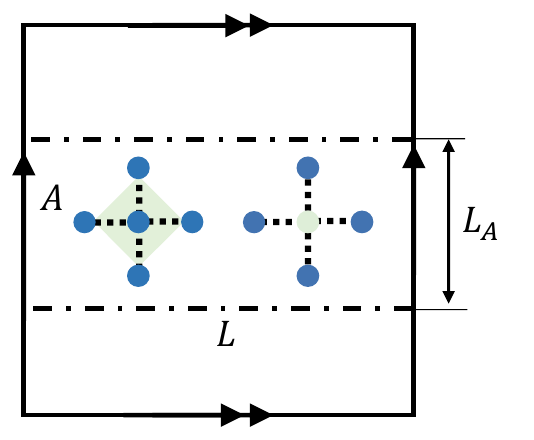}}\qquad
        \subfloat[Shape of subregion $A$ of RTPM on $L\times L$ torus]{\label{fig: ee_set_T}\includegraphics[width=0.22\textwidth]{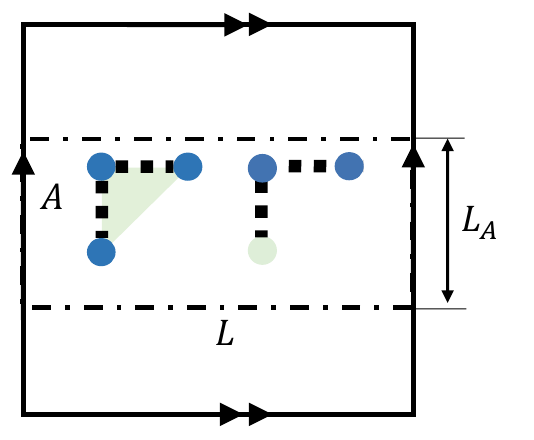}}\qquad
        \subfloat[RXPM critical symmetry entropy scaling $\mathrm{sym}S_A / L$, with $\tilde{c} = 6.9\times 10^{-3}$]{\label{fig: RXPM_ee}\includegraphics[width=0.19\textwidth]{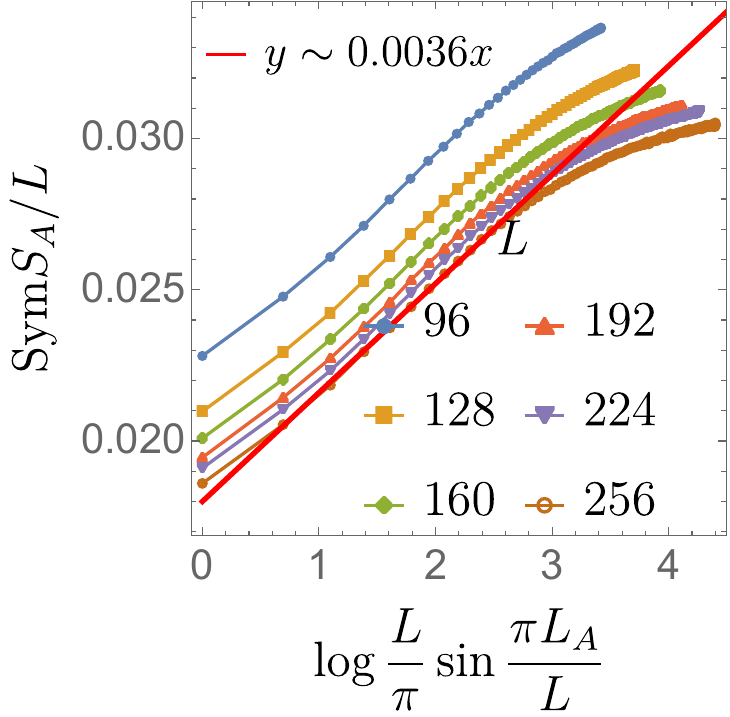}}\qquad
        \subfloat[Trivial scaling of RTPM symmetry entropy]{\label{fig: RTPM_ee}\includegraphics[width=0.19\textwidth]{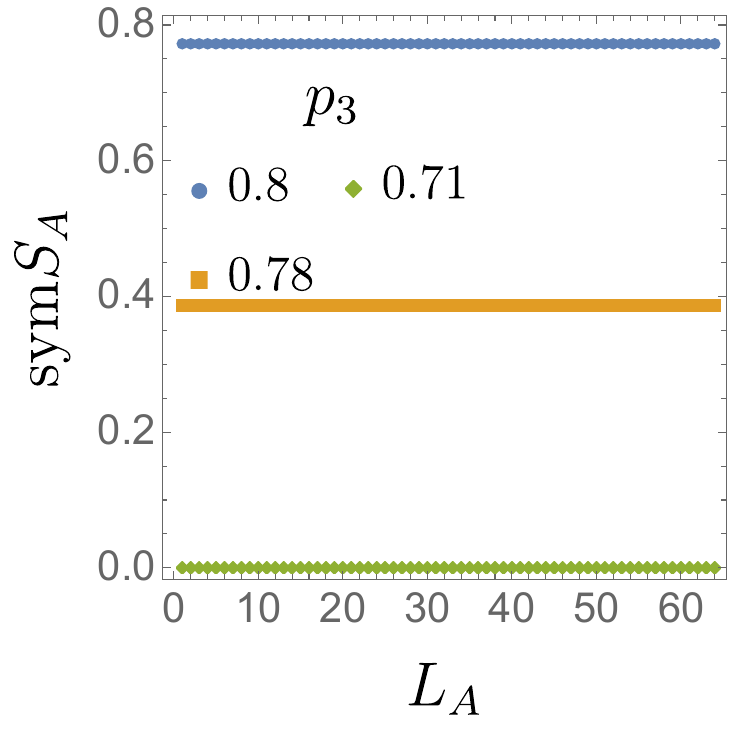}}\qquad
        \subfloat[Half-system symmetry entropy ($L_A = L/2$) of RXPM.]{\label{fig: RXPM_ee_size}\includegraphics[width=0.19\textwidth]{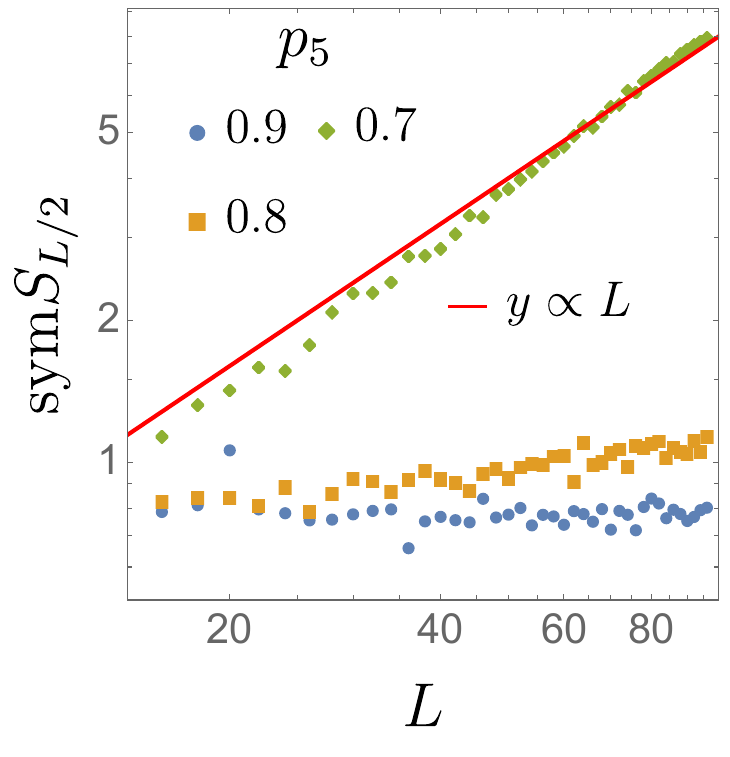}}
        
        \caption{\label{fig: ee}   Data analysis of symmetry entropy $\mathrm{sym}S_A$ of subregion $A$.}
    \end{figure}

    For RXPM model, which is isotropic in both the horizontal and vertical directions, we place this model on an $L \times L$ square torus.  As we tune $p_5$ from $1$ to $0$, a phase transition occurs, as indicated by the configuration entropy density $S_{\text{cf}}/L^2$ shown in Fig.~\ref{fig: XPM_c_entropy}, which suggests that
    \begin{equation}
        S_{\text{cf}} \sim \left\{
            \begin{matrix}
                O(1)   & \quad p_5 > p_5^c\\
                O(L^2) & \quad p_5 < p_5^c
            \end{matrix}
        \right. .
    \end{equation}
When $0<p_5<p_5^c$, the extensive configuration entropy suggests exponential ground state degeneracy in this phase. The phase transition is also captured by the scaling of the symmetry entropy of a subregion $A$:
\begin{equation}
    \mathrm{sym}S_A 
    \sim \left\{
            \begin{matrix}
                O(1) & \quad p_5 > p_5^c\\
                O(l_A) & \quad p_5 < p_5^c
            \end{matrix}
        \right. ,
\end{equation}
where $l_A$ denotes the boundary length of subregion $A$, as present in Fig.~\ref{fig: RXPM_ee_size}. When $p > p_5$, the number of symmetry operators $|G| \sim O(1)$, and the symmetry operator $g\in G$ is non-local, contributing to a constant symmetry entropy of any subregion $A$. As for $p < p_5$, the symmetry operators become local and there are exponential number of them, resulting in an area-law scaling of $\mathrm{sym}S_A$.
At criticality $p_5 \sim 0.743$, we observe that 
\begin{equation}
        \mathrm{sym}S_{A} / L \sim \tilde{c} \log  \frac{L}{\pi} \sin\left(\frac{\pi L_A}{L}\right) + b
\end{equation}
with $\tilde c \sim 3.6 \times 10^{-3}$ and $b$ being some non-universal constant,as shown in Fig.~\ref{fig: ee}.

To further investigate the structure of the symmetry operators, we compute the symmetry mutual information $\mathrm{sym}I_{AB}$ of two disjoint antipodal regions shown in Fig.~\ref{fig: mi_set_p}. As shown in Fig.~\ref{fig: RXPM_mi}, $\mathrm{sym}I_{AB}$ collapses to 
    \begin{equation}\label{eq: bk_nu}
        \mathrm{sym}I_{AB} \sim g(L^{1/\nu_5}(p_5 - p_5^c))
    \end{equation}
    with $\nu_5 = 1.3$ and $p_5^c \sim 0.743$. This finding suggests that when $p_5<p_5^c$, $\mathrm{sym}I_{AB}$ approaches zero in the limit $L\to\infty$, indicating that the symmetry operators are local.

    From the above analysis, it is evident that although both the RTPM and RXPM models exhibit phase transitions, they differ significantly, especially in the phase where $p_{3/5}<p^c_{3/5}$ and at criticality. In the following sections, we will deepen our understanding of the transitions in both models by introducing various boundary conditions.

\section{Boundary physics and cellular automaton dynamics}

In the previous section, we examine the localization phase transition of bulk symmetry operators in RPMs placed on a torus. Here, we shift our focus to models placed on geometries with boundaries. Specifically, we consider RPMs on cylinders with two boundaries, subject to various boundary conditions. Our objective is to study the behavior of symmetry operators originating from the boundary and how they can be used to characterize phase transitions. To achieve this, we introduce a dynamical approach that maps this 2D model to 1D random cellular automaton dynamics. Through this mapping, we demonstrate that the localization phase transition can be equivalently described as a dynamical absorbing phase transition.

\begin{figure}[ht]
        \centering
        \subfloat[\label{fig: open_bd} free boundaries]{\includegraphics[width=0.2\textwidth]{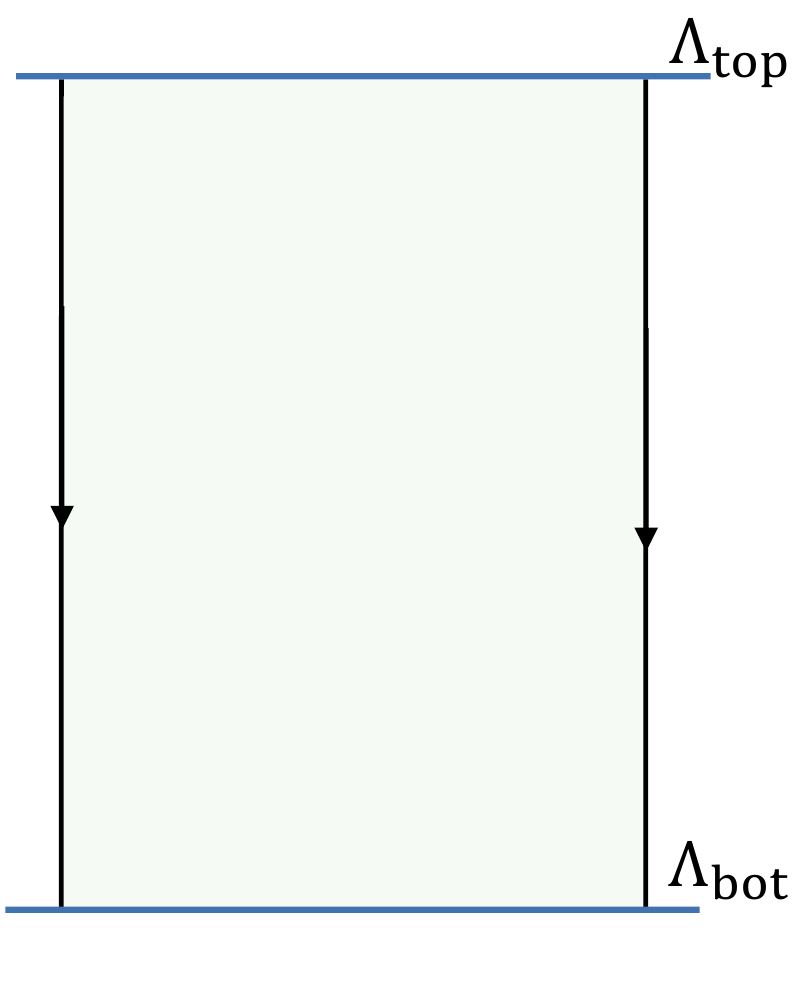}}\qquad
        \subfloat[\label{fig: hf_open_bd} Half-free boundary]{\includegraphics[width=0.22\textwidth]{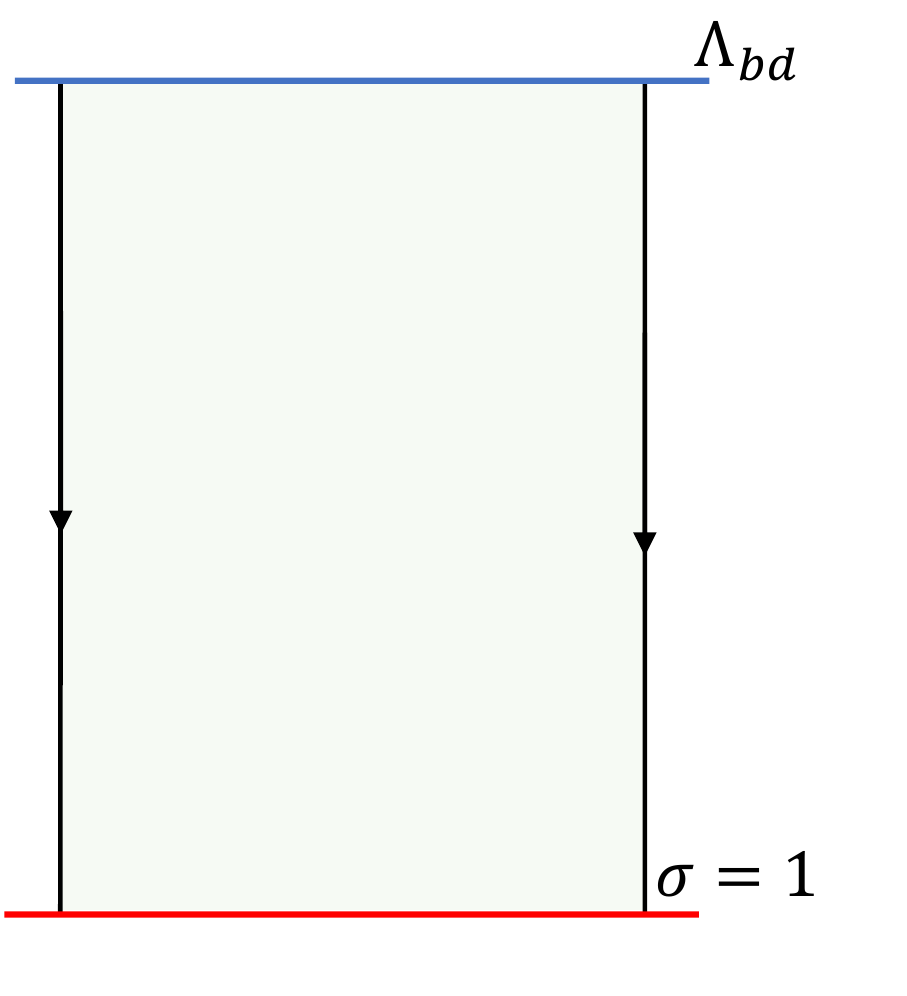}}
        \caption{\label{fig: RPM_bd} RPM defined on cylinder with two boundaries. Black vertical lines with the arrows are glued together. Blue/Red line refers to free/fix boundary respectively. Along the fixed boundary, spins are required to be $\sigma = +1$.}
    \end{figure}

We consider two types of boundaries: fixed and free. In the fixed boundary, all boundary spins $\sigma$ must be $+1$, while in the free boundary, no constraints are imposed on the boundary spins. The free boundary condition effectively removes $L_f$ constraints from the parity check matrix $P$ in Eq.~\ref{eq:const}, resulting in $O(2^{L_f})$ spin-flip symmetries originating from the boundary, where $L_f$ is the length of the free boundary. We consider two combinations of boundary conditions, as illustrated in Fig.~\ref{fig: RPM_bd}: one with only one boundary free and the other fixed, and another with both boundaries free.

We observed a phase transition of the boundary symmetry operator for both the RTPM and RXPM. We demonstrate that the boundary phase transition in RTPM corresponds to the dynamical phase transition of the Domany-Kinzel cellular automaton~\cite{PhysRevLett.53.311, kinzel1985phase}. In contrast, the boundary phase transition in RXPM belongs to the same universality class as the measurement-induced phase transition in the $1+1$D hybrid Clifford circuit~\cite{li2019measurement, skinner2019measurement}. The critical exponents are summarized in Table~\ref{tab: exp1}.

\begin{table}[ht]
    \begin{tabular}{|l|*{6}{p{1cm}|}}
    \hline
    Model       & $z$ & $h^0$ & $h^1$   & $\nu$  & $c$    & $\Delta$ \\ 
    \hline
    RTPM        & $1.697$& $0.692$& $0.075$ & $2.43$ & -  & -     \\
    \hline
    RXPM        & $1$ & $1.53$& $0.125$ & $1.3$  & $3.05$ & $2$      \\  \hline
    \end{tabular}
    \caption{\label{tab: exp1} The critical exponents of RXPM/RTPM boundary criticality are presented. Here, $\nu$ refers to the boundary exponents $\nu_3’$ and $\nu_5$. The boundary exponent $\nu_3’$ is related to the bulk exponent $\nu_3$ (as extracted in the previous section) by the relation $\nu_3’ = z \times \nu_3$.}
\end{table}

\subsection{Spin-flip symmetry and cellular automaton dynamics}\label{sec: auto}

Prior studies have demonstrated that spin-flip symmetry operators of plaquette models can be derived through specific one dimensional classical cellular automata\cite{devakul2019fractal, schuster2023holographic, PhysRevB.88.125122, sfairopoulos2023cellular}. In this part, we first revisit the relationship between deterministic cellular automata (DCA) and plaquette models without randomness. Subsequently, we map the RTPM and RXPM models to two distinct types of randomized cellular automata: the probabilistic cellular automaton (PCA) and cellular automaton with random impurity (CAwRI).

\subsubsection{Deterministic cellular automaton (DCA) dynamcis}

Previously, we have shown in Eq.~\ref{eq: symmtries} that each spin-flip symmetry operator in $G$

    \begin{equation}
        G = \bigg\{g_k = \prod_{i, \tau} X_{i, \tau}^{x_{k,i}^\tau}\bigg\}
    \end{equation}
can be characterized by a binary vector $\vec x_k$ with $k$ running from $1$ to $|G|$. Here $(i,\tau)$ labels a site in the 2D lattice. This can be alternatively understood as a spacetime configuration generated by a cellular automaton at time $\tau$ on lattice site $i$ .

We first start with the Newman-Moore model with 3-body interaction. The constraint imposed in the symmetry operator can be expressed as
    \begin{equation}
        x_{k, i}^{\tau + 1} + x_{k, i}^{\tau} +  x_{k, i + 1}^{\tau} = 0, ~\forall k.
    \end{equation}
    This can be treated as an automata updating rule $x^\tau \to x^{\tau + 1}$
    \begin{equation}
        x_{k, i}^{\tau + 1} =  x_{k, i}^{\tau} +  x_{k, i + 1}^{\tau},
    \end{equation}
    which is the rule-18 automata~\cite{RevModPhys.55.601} that generates the Sierpinski triangle. 
    
    Following similar logic, for the X plaquette model, we have the automata updating rule
    \begin{equation}
        x_{k, i}^{\tau + 1} =  x_{k, i-1}^{\tau} + x_{k, i}^{\tau} +  x_{k, i + 1}^{\tau} +  x_{k, i}^{\tau - 1}.
    \end{equation}
    In this dynamics, the value of $x^{\tau+1}$ is determined by the configurations of the two preceding layers $x^{\tau}$ and $x^{\tau-1}$. This automaton is the reversible second-order automaton shown in~\cite{MARGOLUS198481}.

    \subsubsection{Randomized cellular automata (RCA)}
\begin{figure}[ht]
        \centering
        \subfloat[\label{fig: RTPM_updt}RTPM update rule. 3-body interaction:  $x_{k, i}^{\tau + 1} =  x_{k, i}^{\tau} +  x_{k, i + 1}^{\tau}$; single-site term: $ x_{k, i}^{\tau + 1} = 0$]{\includegraphics[width=0.2\textwidth]{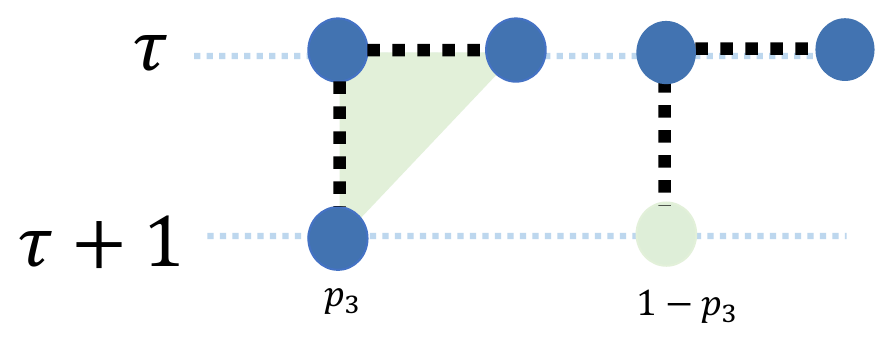}}\qquad
        \subfloat[\label{fig: RXPM_updt}RXPM update rule. 5-body X interaction:  $x_{k, i}^{\tau + 1} =  x_{k, i}^{\tau} +  x_{k, i + 1}^{\tau} +  x_{k, i - 1}^{\tau} +  x_{k, i + 1}^{\tau}$; single-site term: $ x_{k, i}^{\tau + 1} = 0$]{\includegraphics[width=0.2\textwidth]{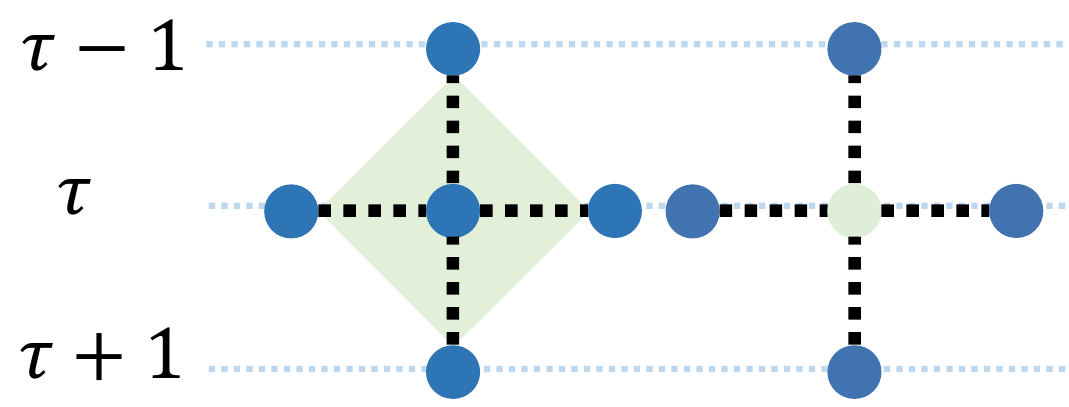}}\qquad
        \caption{\label{fig: RPM_updt} Update rules for RPMs.}
    \end{figure}

    We now discuss the randomized cellular automata dynamics corresponding to RTPM and RXPM. Here we take the spatial/temporal direction shown in Fig~\ref{fig: RPM_updt}.

    \paragraph*{RTPM and Probabilistic cellular automaton (PCA) dynamics} 
    
    For RTPM, if we treat the vertical direction as the time direction shown in Fig.~\ref{fig: RTPM_updt}, it can be mapped to a 1+1 dimensional classical cellular automata dynamics. The update rule is defined as
   \begin{equation}
        \begin{tabular}{c | l}\label{eq: RTPM_updt}
            $\mu (1 | 00) = 0$ & $\mu (0 | 00) = 1$\\
            $\mu (1 | 10) = p_3$& $\mu (0 | 10) = 1-p_3$\\
            $\mu (1 | 01) = p_3$& $\mu (0 | 01) = 1-p_3$\\
            $\mu (1 | 11) = 0$& $\mu (0 | 11) = 1$\\
        \end{tabular}
    \end{equation}
    where $\mu(x_i^{\tau + 1} | x_i^{\tau} x_{i + 1}^{\tau})$ denotes the probability of site $i$ taking value $T_i^{\tau + 1}$ at time $\tau +1$ under the condition that $x_i^{\tau}$ and $x_{i + 1}^{\tau}$ takes some certain value at time $\tau$. This update rule is the same as the one in Wolfram rule 102 and is a special limit of the Domany-Kinzel model~\cite{PhysRevLett.53.311, kinzel1985phase, A_Kemper_2001}. We notice that $\mu(0 | 0 0 ) = 1$, effectively inhibiting the emergence of the symmetry operator originating from the bulk.
    
    \paragraph*{RXPM and cellular automaton with random impurity (CAwRI)}
    
    For RXPM, we again take the vertical direction as the time direction shown in Fig.~\ref{fig: RXPM_updt}. When the random interaction takes the five-body form, the update rule is
    \begin{equation}
        x_{k, i}^{\tau + 1} =  x_{k, i}^{\tau} +  x_{k, i + 1}^{\tau} +  x_{k, i - 1}^{\tau} +  x_{k, i}^{\tau - 1}.
        \label{eq:5-body}
    \end{equation}
    The single-site term can be viewed as an impurity in the automaton dynamics, imposing a constraint on the $T$ matrix, i.e.,
    \begin{equation}
        x^{\tau}_{k, i} = 0, ~\forall (i,\tau)\in \Lambda_I,
    \end{equation}
    where $\Lambda_I$ are these impurity sites. Furthermore, this constraint leads to the generation of a new symmetry operator, starting at $(i, \tau + 1)$.

\begin{figure}[ht]
         \centering
         \subfloat[\label{fig: RXPM_phy_0} The dynamics up to the $(\tau-1)$th layer]{\includegraphics[width=0.2\textwidth]{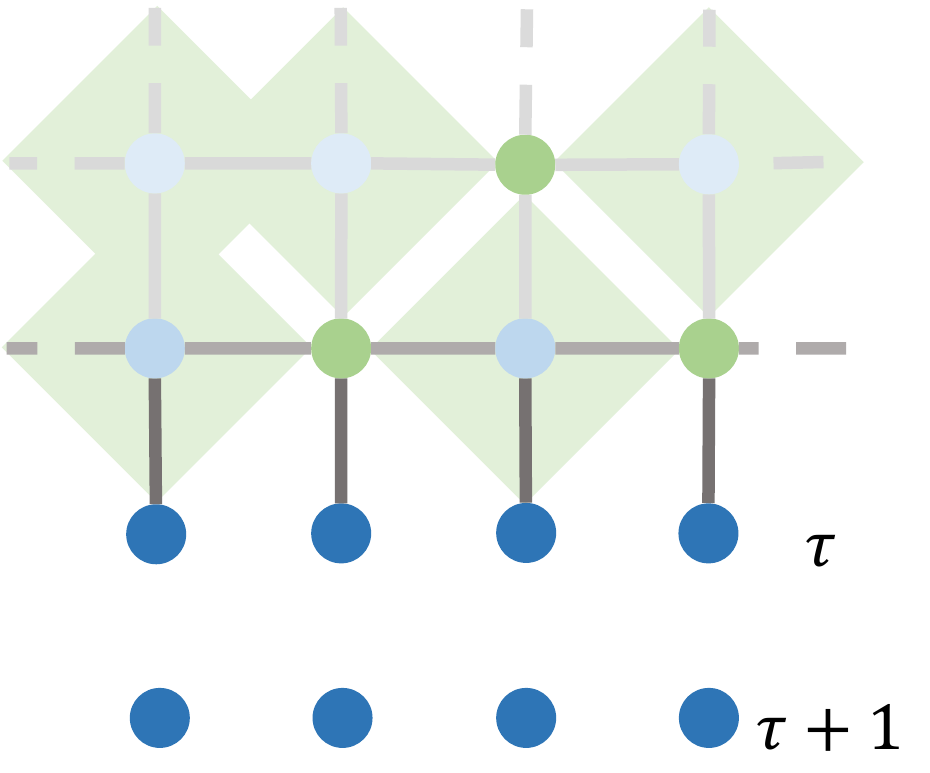}}\qquad
         \subfloat[\label{fig: RXPM_phy_1} The dynamics at the $\tau$th layer. The on-site term, denoted by the green dot on the $\tau$th layer, induces a single site spin-flip $X$ on the $(\tau+1)$th layer denoted as the yellow dot.]{\includegraphics[width=0.2\textwidth]{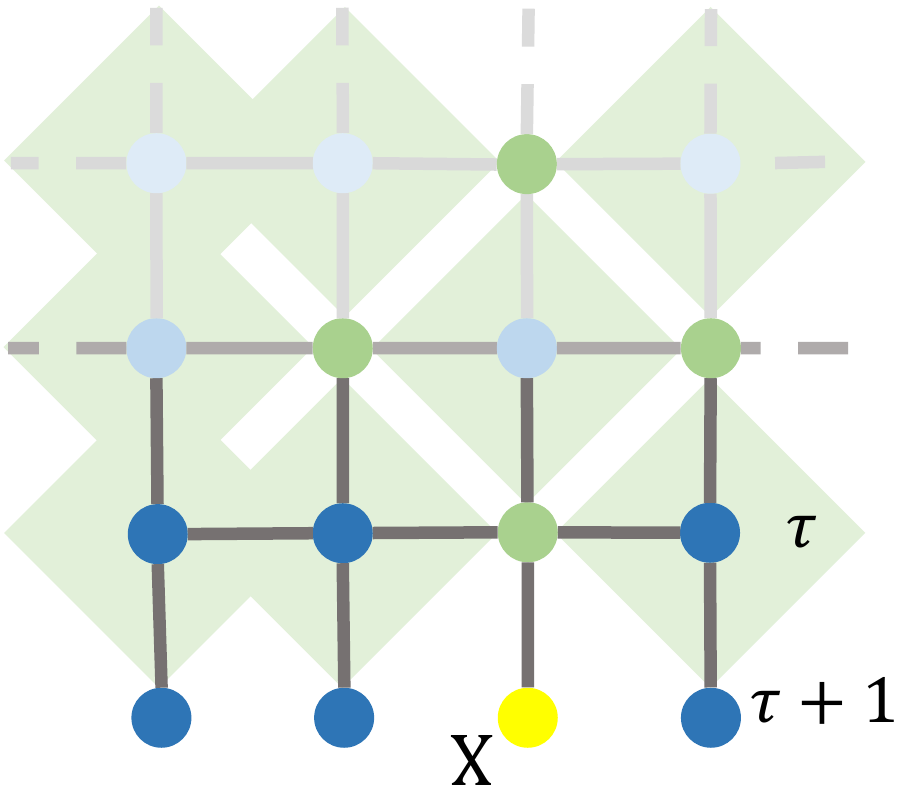}}\qquad
         \caption{\label{fig: RXPM_evo} The cellular automaton dynamics of the RXPM model.}
     \end{figure}

To find the symmetry operators satisfying the above constraint, we propose an algorithm to compute all generators of the symmetry group. Here's the approach: Assuming we have already obtained the symmetry group $G^\tau$ at $t=\tau$ generated by $g_k^{\tau}$. We now want to add one more layer as shown in Fig.~\ref{fig: RXPM_evo} and compute the generators $g_k^{\tau+1}$ of $G^{\tau+1}$. For the five-body interaction centered at $t=\tau$, $x_{k,i}^{\tau+1}$ follows the update rules defined in Eq.\eqref{eq:5-body}. However, if we apply an on-site $\sigma_{i,\tau}$, one requires $x^{\tau}_{k, i} = 0$.

We denote generators that satisfy this constraint as $\Omega^1 = \{g^1_m\}$ and those that fail to meet this constraint as $\Omega^2 = \{g^2_n\}$. For the generators that do not meet this constraint, we select the first one $g_1^2\in \Omega^2$ from the unsatisfied set and modify the remaining generators to 
\begin{equation}
    \tilde{g}^2_n=g_1^2g_n^2
\end{equation}
 so that the associated $\tilde{x}^{\tau}_{n, i} = 0$.
 We then remove $g_1^2$ from the $G^\tau$. Moreover, the on-site interaction induces a free spin at $t=\tau + 1$, introducing an extra spin flip symmetry $X_i$. The group generators for the $\tau + 1$ layer system are now
    \begin{equation}
        G^{\tau + 1} = \langle g^1_m,~ \tilde{g}^2_n, ~X_{i,\tau+1}\rangle.
    \end{equation}
Note that the generators constructed in this manner may not be independent. To resolve this, at the end of the evolution, it is necessary to apply Gaussian elimination to eliminate any redundancies.
In this algorithm, the initial condition at $t=1$ is chosen as 
\begin{align}
    G^{1}=\langle X_{i,1},  X_{i, 2}\rangle
\end{align}
with $i=1,\cdots L$. Here two layers of free spins are needed to evaluate the third layer.

\subsection{Dynamical phase transitions and the boundary physics}

Due to the existence of the boundary, the symmetries $g \in G$ are now categorized into two types: those with nontrivial support on the boundary and those supporting trivially on the boundary. We define symmetries $g \in G$ supporting trivially on the boundary as the bulk symmetry $g^{bk}$. The bulk symmetries form a group $G^{bk} \subset G$. In Sec.~\ref{sec: sym}, we have analyzed the transitions in these bulk operators. In this section, we are interested in these operators with non-trivial support on the boundary and we study the transitions of these operators. They form a group and can be obtained as the quotient
\begin{equation}\label{eq: g_bd}
        G^{bd} \equiv G / G^{bk}.
\end{equation}
The size of $G^{bd}$ can be obtained by
\begin{equation}
    \mathrm{rank} T^{bd} = \mathrm{rank} T - \log|G^{bk}| = \log|G^{bd}|,
\end{equation}
following the same logic as in Eq.~\ref{eq: rank_nullity}. One may regard $G^{bk}$ as the disconnected contribution to the boundary. Here the submatrix $T^{bd}$ is obtained by directly truncating the full tableau $T$ in Eq.~\ref{eq: t_mat}, taking only entries associated with the boundary.

For the cylinders with free boundary condition on both ends, we are interested in these boundary operators which can have non-trivial support on both boundaries. This form a subgroup of $G^{bd}$ and the number of symmetry generators of this group is now given by
\begin{equation}\label{eq: puri_mi}
\begin{aligned}
     \mathrm{sym}I &= \log\left| G^{bd} / \left(G_{\text{top}}^{bd} G_{\text{bot}}^{bd}\right) \right|\\
     &=  \mathrm{rank} T_{\text{top}}^{bd} + \mathrm{rank} T_{\text{bot}}^{bd} - \mathrm{rank} T^{bd},
\end{aligned}
\end{equation}
where $T_{\mathrm{top/bot}}$ represents the submatrix of $T$ taking only entries related to the top or bottom boundaries. From the viewpoint of cellular automaton dynamics, the matrices $T_{\mathrm{top/bot}}$ are now the initial/final state tableau $T_{1/\tau}$ of the cellular automaton dynamics.

The definition of boundary symmetry $G^{bd}$ also extends to the cylinder with one fixed boundary shown in Fig.~\ref{fig: hf_open_bd}, where the boundary symmetry operators only start on the free boundary. These boundary symmetry operators can also be used to characterize the phase transition. Analogous to the bulk symmetry operator, we now define the boundary symmetry entropy as
\begin{equation}\label{eq: Sbd}
      \begin{aligned}
          &\mathrm{sym}S_A^{bd} \equiv\log\left| G^{bd} / \left(G_A^{bd} G_{\overline{A}}^{bd}\right) \right|\\
          =& \mathrm{rank} T_{A}^{bd} + \mathrm{rank} T_{\overline{A}}^{bd} - \mathrm{rank} T^{bd},
      \end{aligned}
\end{equation}
which quantifies the number of group generators that have both nontrivial support on the boundary domain $A$ and its boundary complement $\overline{A}$. We can further define the boundary mutual information as
\begin{equation}
    \mathrm{sym}I_{AB}^{bd} =  \mathrm{sym}S_A^{bd} +  \mathrm{sym}S_B^{bd} -  \mathrm{sym}S_{AB}^{bd}
\end{equation}
for two subsystems $A$ and $B$ on the boundary.

In what follows, we investigate the boundary phase transition using the symmetry measures defined here. Through numerical analysis, we demonstrate a transition from an active phase where boundary operators can penetrate deeply into the bulk and exhibit extensive support there, to an absorbing phase where boundary operators become localized near the boundary, as we vary the parameter $p_{3/5}$ in the RPMs. Remarkably, the critical point $p_{3/5}^{bc}$ closely aligns with the one identified in the bulk localization transition, suggesting that these transitions are indeed identical.

\subsubsection{ Free boundaries on both sides}

We begin by initializing a symmetry operator at the top boundary and analyze its evolution over time. 

As we increase the probability of the single-site term, the fraction of zeros in this operator becomes more prevalent. Eventually, when $p<p^c$, the entire operator rapidly decays to zero within the bulk of the system, and becomes localized near the top boundary. We refer to this as the absorbing phase.

In both models, we observe that for $p>p^c$, the operators can rapidly spread throughout the entire system due to the update rule determined by the three-body or five-body interaction. The occasional single-site term, which enforces the corresponding site to remain zero, is unable to hinder the operator's growth. Even after a prolonged time evolution, the operator can reach the bottom boundary. We refer to this as the active phase.

From this perspective, it is natural to introduce $\mathrm{sym}I$, as defined in Eq.~\eqref{eq: puri_mi}, between the top and bottom boundaries to characterize this absorbing phase transition. Our numerical findings indicate that while both models undergo phase transitions, they belong to distinct universality classes.

    \begin{figure}[ht]
         \centering
         \subfloat[\label{fig: DKCA0.9o} $p_3 = 0.9$]{\includegraphics[width=0.12\textwidth]{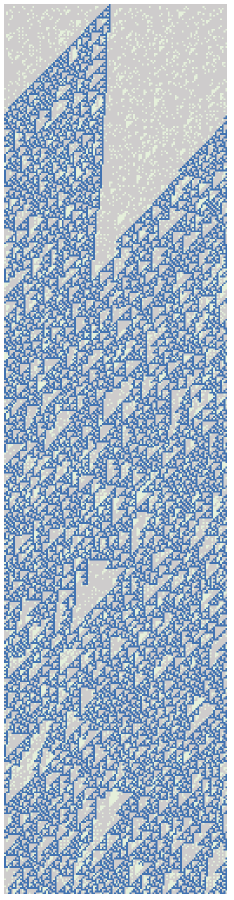}}\qquad
         \subfloat[\label{fig: DKCA0.81o} $p_3 = 0.81$]{\includegraphics[width=0.12\textwidth]{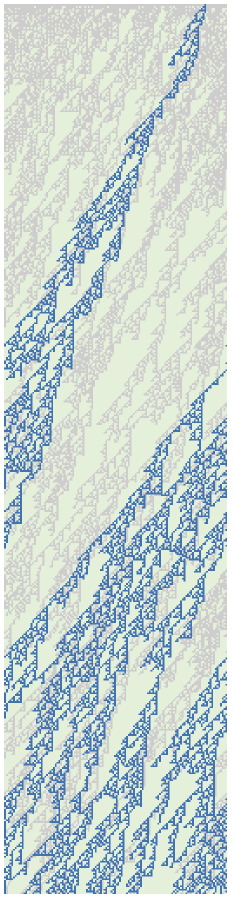}}\qquad
         \subfloat[\label{fig: DKCA0.75o} $p_3 = 0.75$]{\includegraphics[width=0.12\textwidth]{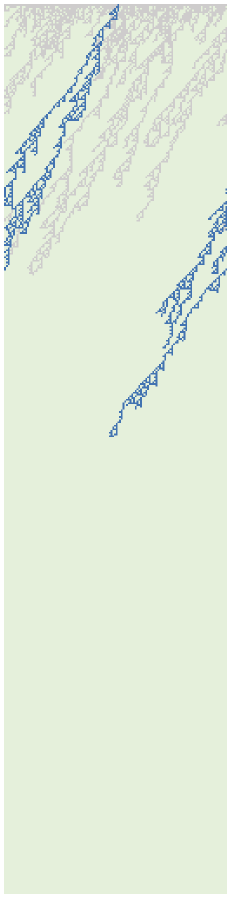}}\qquad
         \caption{\label{fig: RTPMo_fig} Space-time configuration of the RTPM-PCA evolution. Time flows from top to bottom. Each blue site hosts a spin-flip operator. Gray sites label the support of the full boundary group $G^{bd}$.
         }
     \end{figure}

\paragraph{RTPM effective dynamical transition}

\begin{figure}[ht]
        \centering
        \subfloat[$\mathrm{sym}I$ of RTPM.]{\label{fig: TPM_bd_mi2}\includegraphics[width=0.3\textwidth]{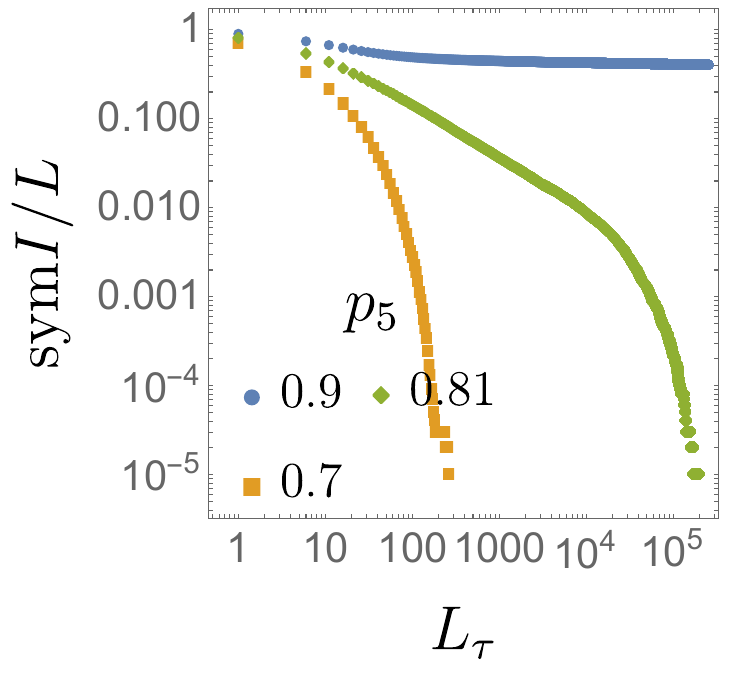}}\qquad
        \subfloat[critical scaling of $\mathrm{sym}I$ of RTPM.]{\label{fig: RTPM_bd_mi_crit2}\includegraphics[width=0.3\textwidth]{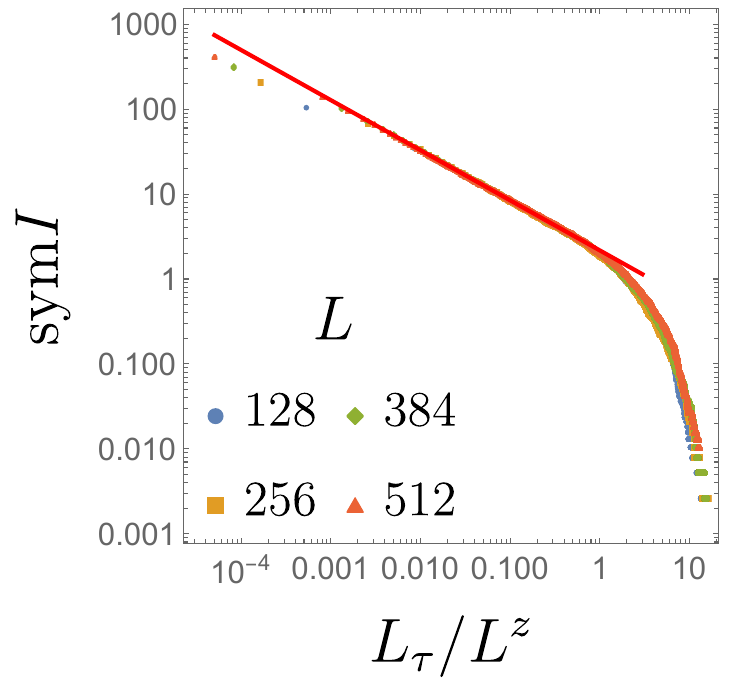}}
        \caption{\label{fig: sym_bd_mi2_T} $\mathrm{sym}I$ between top and bottom boundaries.}
    \end{figure}

The scaling of $\mathrm{sym}I$ with the effective time $L_{\tau}$ and boundary length $L$ captures different phases of the RTPM. As shown in Fig.~\ref{fig: TPM_bd_mi2}, 
\begin{equation}
    \mathrm{sym}I \sim
    \begin{cases}
        L & p_3 > p_3^c\\
        L \exp(- d L_\tau) & p_3 < p_3^c
    \end{cases}
\end{equation}
where $d > 0$ is some non-universal number, and $p_3^c = 0.81$.  When $p_3 > p_3^c$, the update rule corresponding to the three-body interaction dominates. In this regime, symmetry operators originating from one boundary can propagate to the other boundary even when $L_\tau\gg L$ (Fig.~\ref{fig: DKCA0.9o}). This behavior leads to a plateau observed in Fig.~\ref{fig: TPM_bd_mi2}. On the other hand, for $p_3 < p_3^c$, the frequent single-site forces the symmetry operator to take trivial values and therefore the operator starting from the boundary quickly vanishes as shown in Fig.~\ref{fig: DKCA0.75o}. This results in an exponential decay in $\mathrm{sym}I$, and thus the two boundaries are not connected by the boundary symmetry operators when $L_\tau \gg L$.

The critical behavior of $\mathrm{sym}I$ is presented in Fig.~\ref{fig: RTPM_bd_mi_crit2}, where $\mathrm{sym}I$ collapses to a universal scaling function:
\begin{equation}
    \mathrm{sym}I \sim h(L_\tau/L^z),
\end{equation}
where the scaling function takes the form
\begin{equation}
    h(\theta_\tau) \sim \left\{
    \begin{matrix}
         \pi h^0 \theta_\tau^{-1/z} & \theta_\tau \ll 1\\
         \exp(- h^1 2\pi \theta_\tau ) &\theta_\tau \gg 1
    \end{matrix}\right.
\end{equation}
with dynamical exponent $z = 1.697$, early time exponent $h^0 = 0.692$, and late time exponent $h^1 = 0.075$. The dynamical exponent $z$ is very close to the one computed in ~\cite{PhysRevE.66.016113}.

\begin{figure}[ht]
        \centering
        \includegraphics[width=0.3\textwidth]{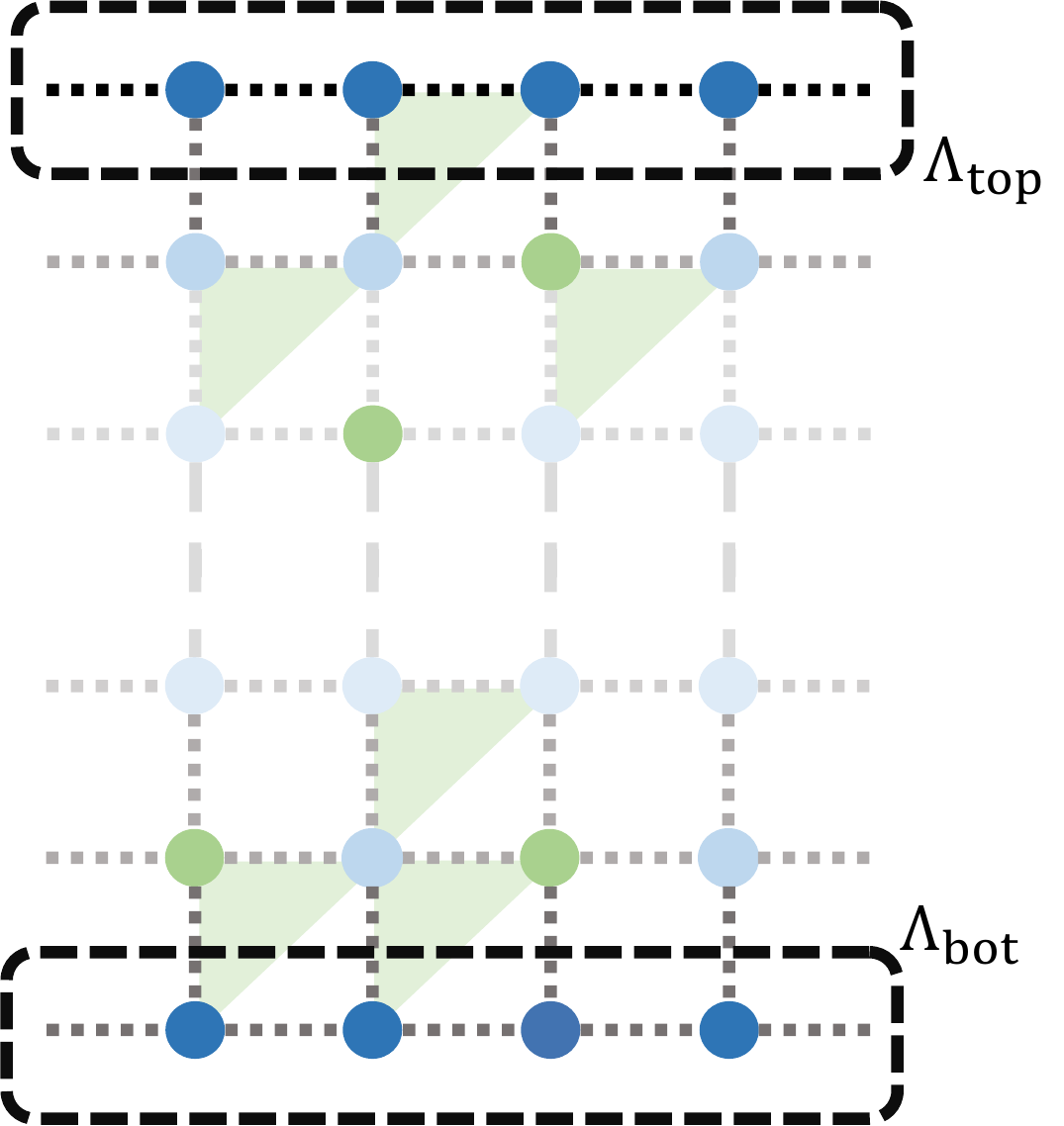}
        \caption{ \label{fig: RTPM_bd} RTPM with free boundaries (dark blue dots) on top and bottom layers. The bulk three-body couplings are illustrated as green triangles and the single-site terms are illustrated as the green dots.}
           
\end{figure}

    \begin{figure}[ht]
         \centering
         \subfloat[\label{fig: CAwRIo0.8} $p_5 = 0.8$]{\includegraphics[width=0.12\textwidth]{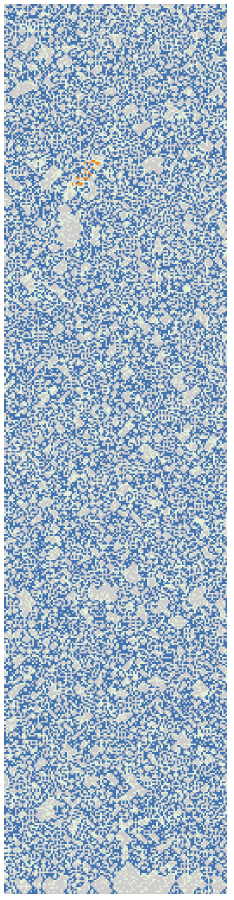}}\qquad
         \subfloat[\label{fig: CAwRIo0.743} $p_5 = 0.743$]{\includegraphics[width=0.12\textwidth]{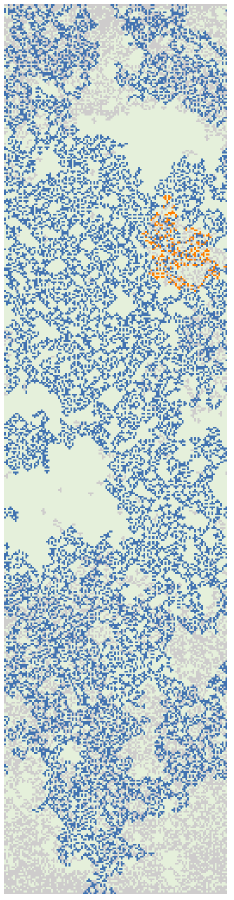}}\qquad
         \subfloat[\label{fig: CAwRIo0.7} $p_5 = 0.7$]{\includegraphics[width=0.12\textwidth]{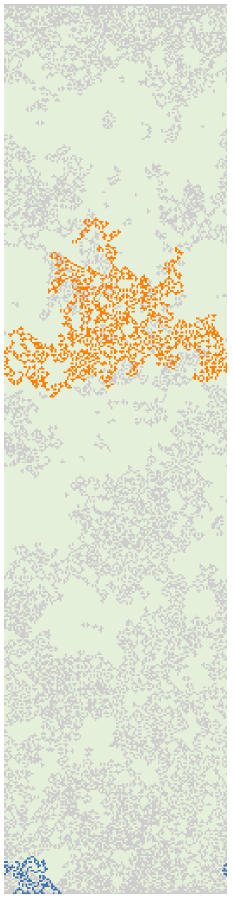}}\qquad
         \caption{\label{fig: RXPM_physics} Space-time configuration of the RXPM-CAwRI evolution. Time flows from top to bottom. Blue sites host spin-flips symmtries associated with the boundary, the orange sites carry spin-flips related to the bulk, and the grey sites support the full symmetry group.}
    \end{figure}

\paragraph{RXPM effective dynamical transition}

\begin{figure}[!h]
            \centering
            \subfloat[$\mathrm{sym}I$ of RXPM]{\label{fig: XPM_bd_mi2}\includegraphics[width=0.34\textwidth]{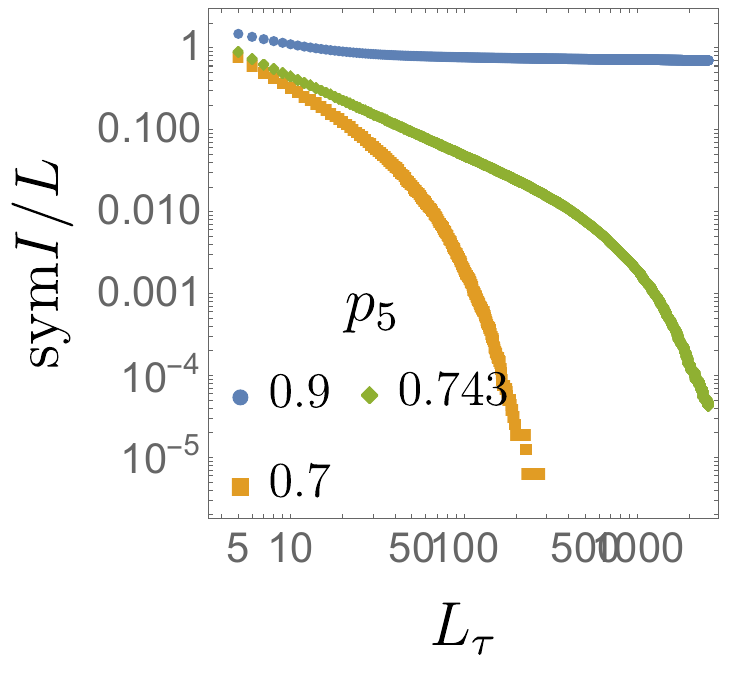}}\qquad
            \subfloat[Early time critical scaling of $\mathrm{sym}I$ of RXPM. Red line: $y = 1.53 (L_{\tau}/L)^{-1}$]{\label{fig: RXPM_bd_mi_crit2}\includegraphics[width=0.34\textwidth]{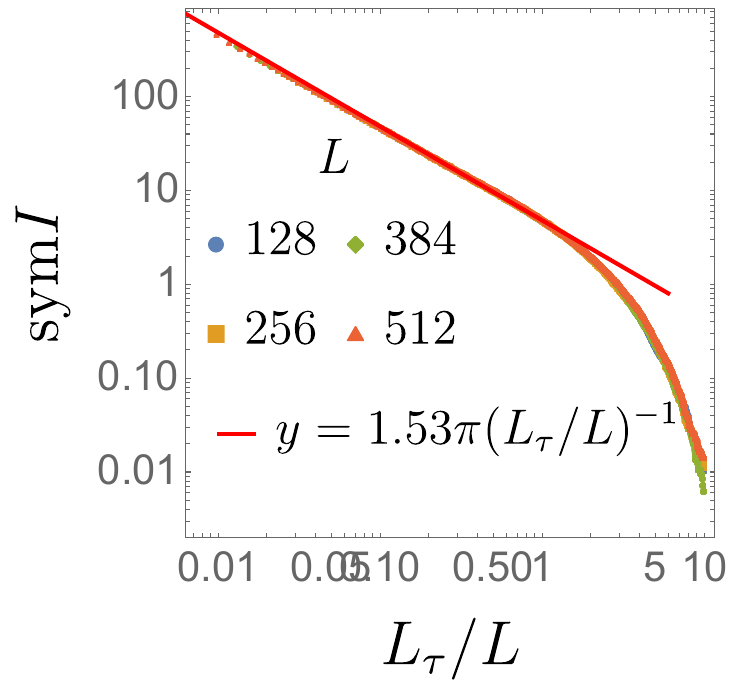}}\qquad
            \subfloat[Late time critical scaling of $\mathrm{sym}I$ of RXPM. Red line: $y \propto \exp(-0.125 \times 2\pi L_\tau / L)$]{\label{fig: XPM_bd_mi_dc2}\includegraphics[width=0.34\textwidth]{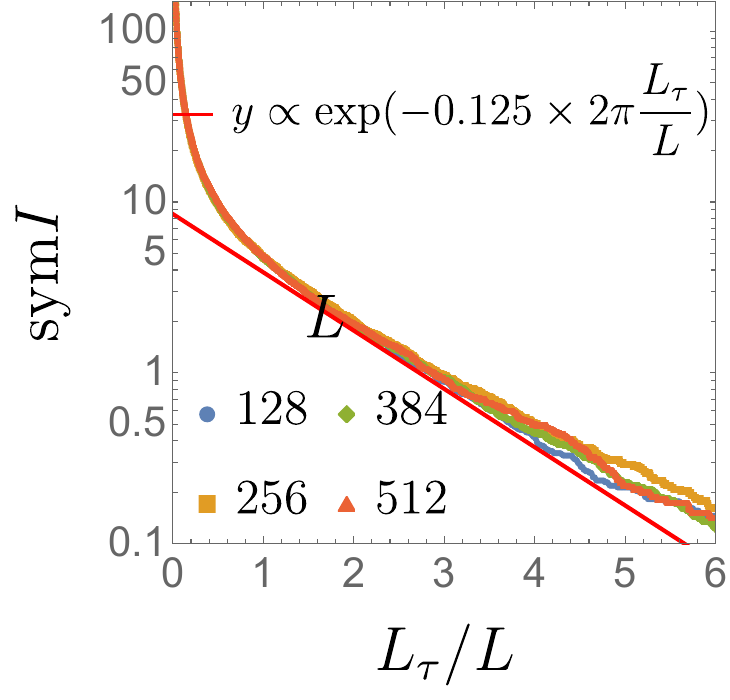}}
            \caption{\label{fig: sym_bd_mi2_X} Symmetry mutual information between the top boundary and the bottom boundaries.}
    \end{figure}

For RXPM, when the five-body coupling terms dominate, the symmetries are non-local, and the top and bottom boundaries are connected by non-trivial symmetry generators as shown in Fig.~\ref{fig: CAwRIo0.8}. In contrast, in the area-law phase, the symmetry operators are local, resulting in an exponential decay in $\mathrm{sym}I$. The scaling of mutual information $\mathrm{sym}I$ between the top and bottom boundary is again
    \begin{equation}
        \mathrm{sym}I\sim\left\{
        \begin{matrix}
            L & p_5 > p_5^c\\
            L \exp(- d L_{\tau}) & p_5 < p_5^c
        \end{matrix}
        \right. .
    \end{equation}
with $d > 0$ being some non-universal exponent.
As shown in Fig.~\ref{fig: sym_bd_mi2_X}, at criticality, $\mathrm{sym}I$ collapse to a universal scaling function
\begin{equation}
    \mathrm{sym}I \sim h(L_\tau / L),
\end{equation}
and the critical scaling function takes the form
    \begin{equation}
        h(\theta_\tau) \sim \left\{
            \begin{matrix}
              \pi h^0\theta_\tau^{-1}& \theta_\tau \ll 1\\
                \exp (-h^1  2 \pi \theta_\tau) & \theta_\tau \gg 1
            \end{matrix}
            \right.
\end{equation}
Here, we have early time exponent $h^0 = 1.53$, and late time exponent $h^1 = 0.125$.
    \begin{figure}[ht]
        \centering
        \includegraphics[width=0.3\textwidth]{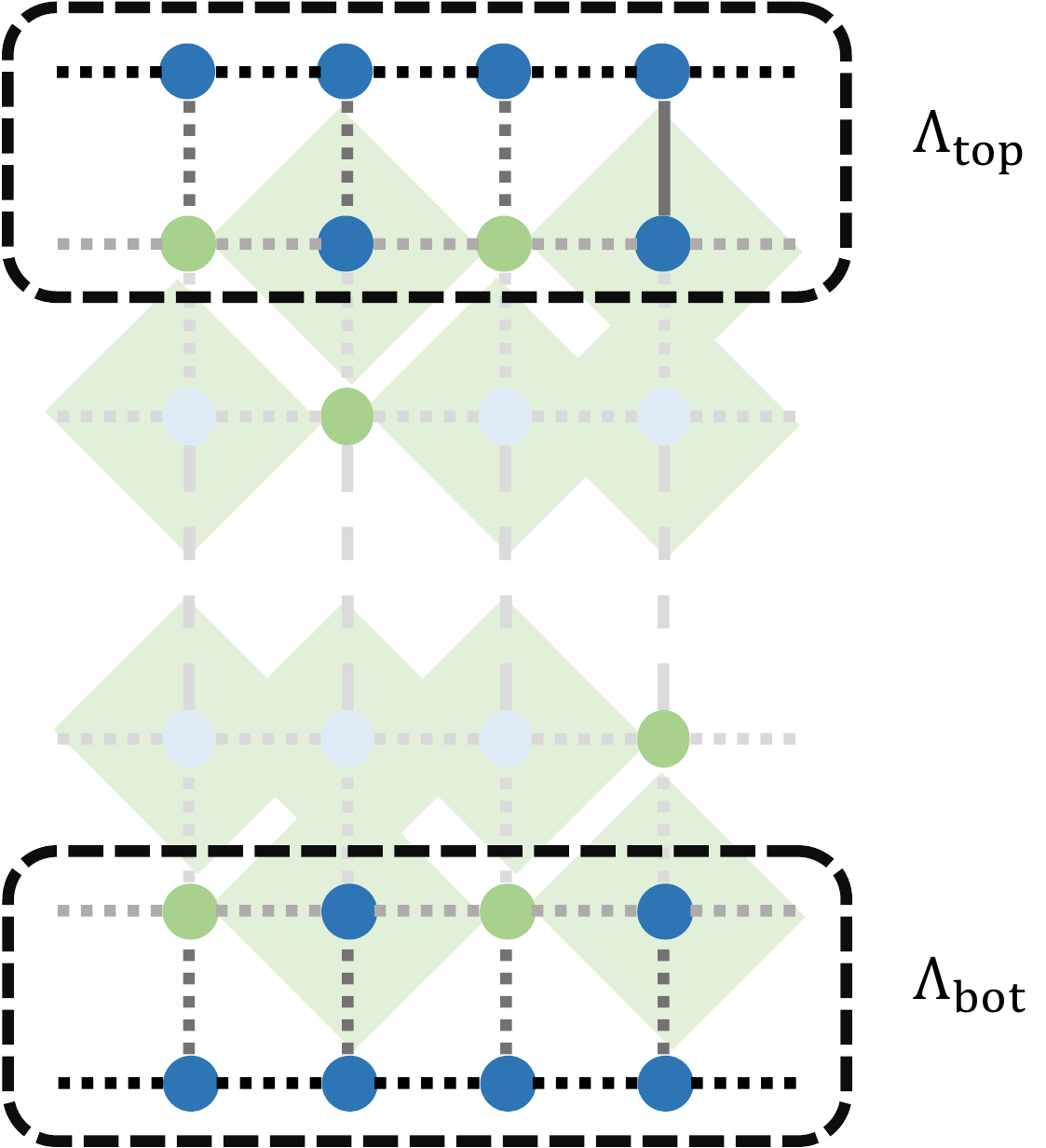}
            \caption{RXPM with free boundaries (dark blue dots) on top and bottom layers. The bulk five-body interactions are illustrated as green squares and the single-site terms are illustrated as green dots.}
            \label{fig: RXPM_open_bd}
    \end{figure}

\subsubsection{ One boundary free and the other fixed}

The boundary phase transition is also captured by the boundary symmetry operator of a symmetry with half-free boundary condition, as shown in Fig.~\ref{fig: hf_open_bd}. In this case, the boundary symmetry only relates to the free boundary. 

\begin{figure}[ht]
        \centering
        \includegraphics[width=0.3\textwidth]{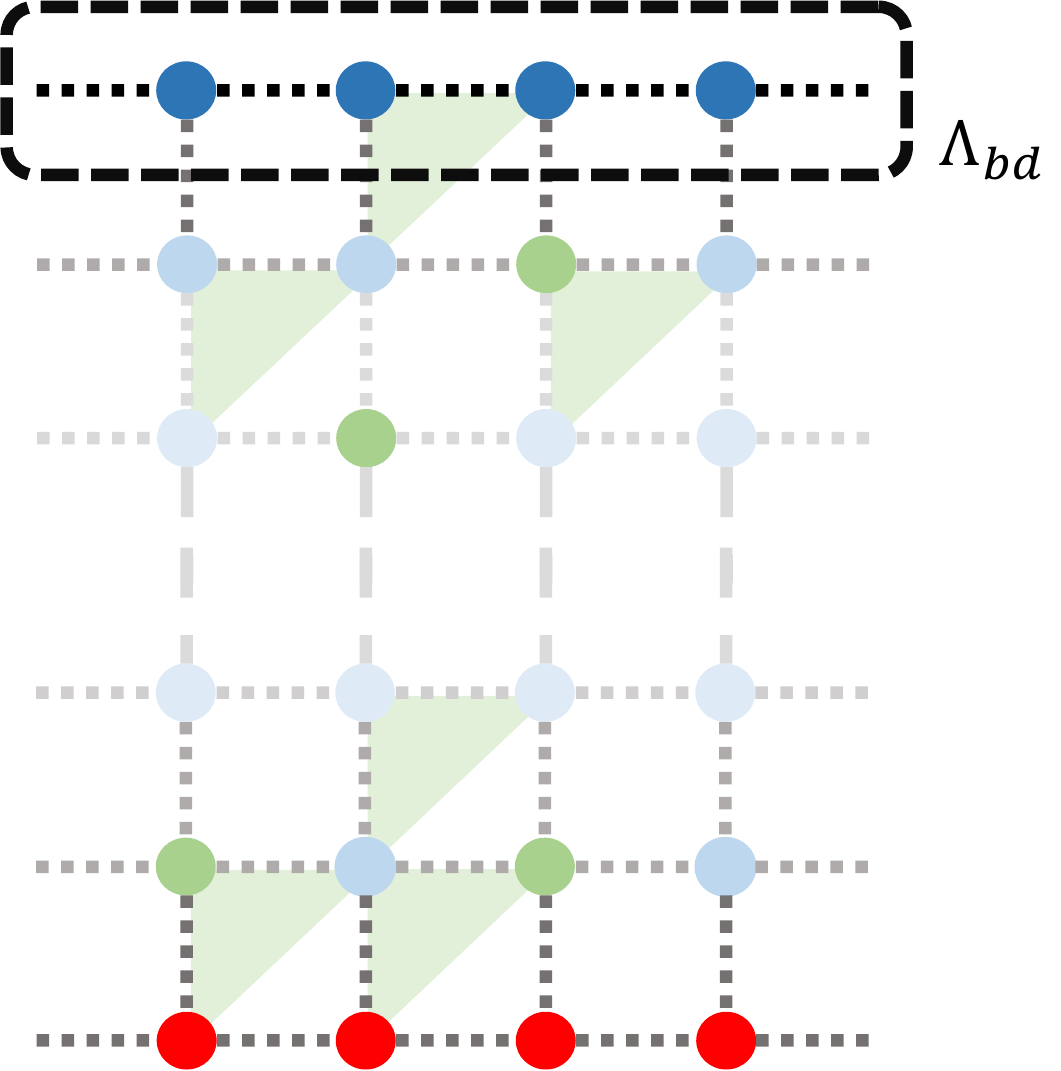}
            \caption{RTPM with free boundary on top (dark blue dots) and fixed boundary on bottom (red dots). The bulk three-body interactions are illustrated as green triangles and the single-site terms are illustrated as green dots.}
            \label{fig: RTPM_bd_hf_open}
    \end{figure}

\paragraph{RTPM boundary transition}
We investigate the boundary phase transitions in RTPM with half-free boundary as shown in Fig.~\ref{fig: RTPM_bd_hf_open}. The symmetry operators are evaluated via the randomized updating rule presented in Eq.~\ref{eq: RTPM_updt}. The fixed boundary condition is realized by imposing constraints on the state-vectors in the following manner.

We encode the initial conditions of the RTPM automaton into a $L \times L$ matrix $T^0$, where each row vector in matrix $T^0$ represents an initial condition. For simplicity, one can take $T^0 = I_L$, the identity matrix, where each initial condition only contains one non-trivial entry. Evolving $T^0$ $L_\tau$ times, we obtain the final state matrix $T^\tau$, where each row vector is related to the initial state vector. The full boundary tableau
\begin{equation}\label{eq: dyno_tab}
    T^{bd} = \left(T^0 \mid T^\tau\right)
\end{equation}
comprises two parts, the initial state tableau $T^0$ and the final state tableau $T^\tau$.

The fixed boundary condition requires that the generator must satisfy $x^\tau = 0$ at the final time. When $p_3 < p_3^{bd}$, if $L_\tau$ is large, the generator rapidly decays to zero in the bulk, automatically satisfying $T^\tau = 0$. However, when $p_3 > p_3^{bd}$, the symmetry operator starting from one boundary can reach the other boundary even after the long time evolution. To identify the subspace of symmetry operators for which $x^\tau=0$, we use the standard row reduction techniques. After the row reduction, the state matrix takes the form
\begin{equation}\label{eq: bd_fix}
   T^{bd} = \left( 
   \begin{matrix}
       Q^0 & Q^\tau \\
       Q^1 & 0
   \end{matrix}\right),
\end{equation}

    \begin{figure}[ht]
         \centering
         \subfloat[\label{fig: DKCA0.9} $p_3 = 0.9$]{\includegraphics[width=0.12\textwidth]{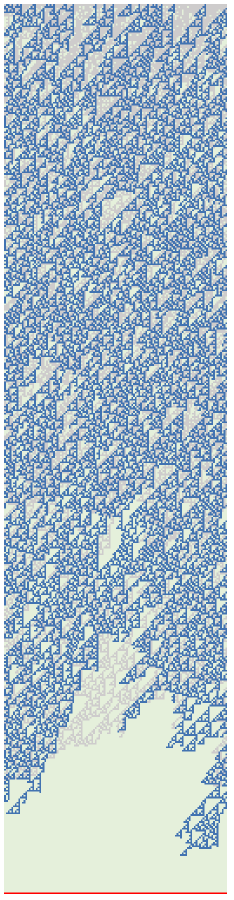}}\qquad
         \subfloat[\label{fig: DKCA0.81} $p_3 = 0.81$]{\includegraphics[width=0.12\textwidth]{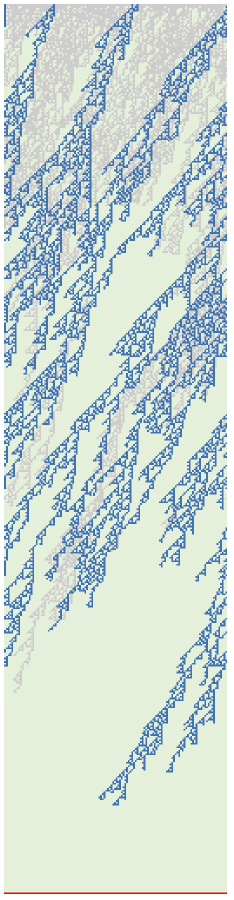}}\qquad
         \subfloat[\label{fig: DKCA0.75} $p_3 = 0.75$]{\includegraphics[width=0.12\textwidth]{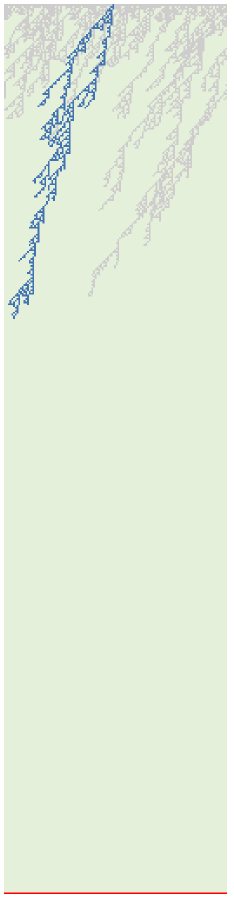}}\qquad
         \caption{\label{fig: RTPM_physics} Space-time configuration of the RTPM-PCA evolution. Time flows from top to bottom. The bottom boundary is fixed, labeled by red. Each blue site hosts a spin-flip operator, and the initial condition is a single spin-flip $X$ reside in the middle. Grey sites label the support of the full boundary group $G^{bd}$.
         }
     \end{figure}

\begin{figure*}[ht]
        \centering
        \subfloat[RTPM boundary symmetry mutual information data collapse]{\label{fig: RTPM_bd_mi_dc}\includegraphics[width=0.28\textwidth]{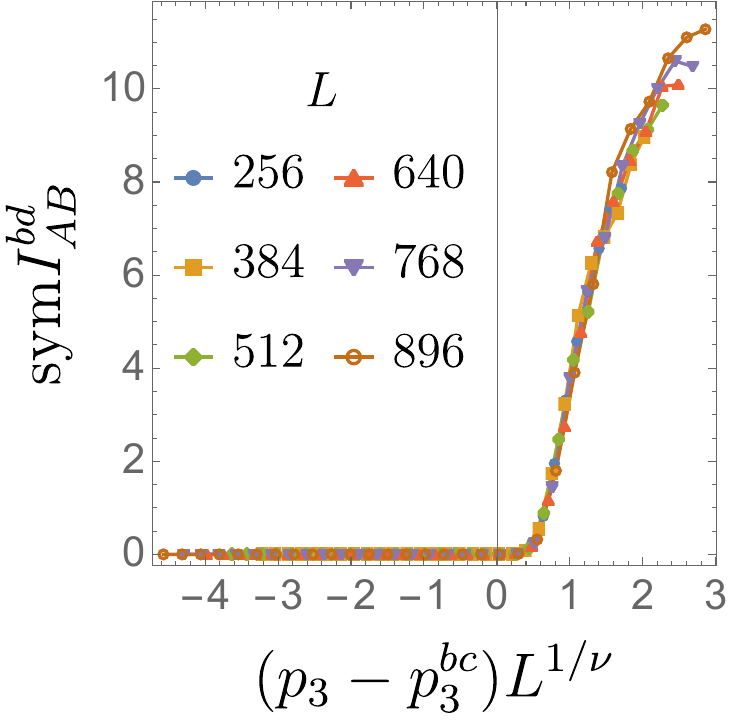}}\qquad
         \subfloat[Geometry of the boundary symmetry mutual information $A$ and $B$, with $L_A = L_B = L / 8$ ]{\label{fig: mi_bd_set}\includegraphics[width=0.15\textwidth]{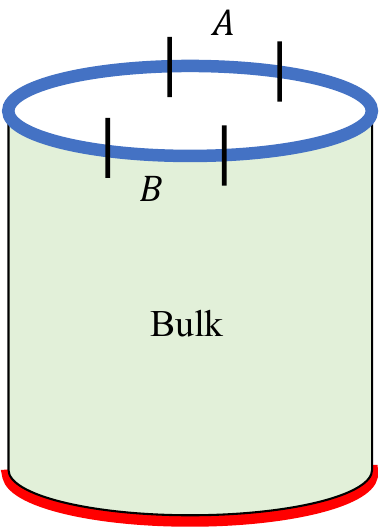}}\qquad
        \subfloat[RTPM boundary symmetry entropy scaling]{\label{fig: RTPM_bd_ee_L}\includegraphics[width=0.28\textwidth]{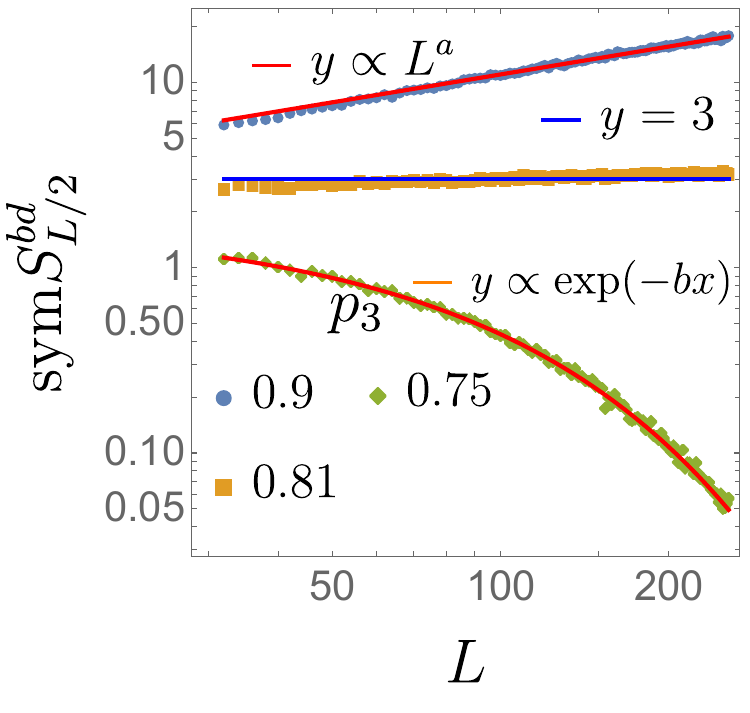}}\qquad
        \subfloat[Geometry of half entropy $\mathrm{sym}S_{L/2}$. Here $L_A = L/2$]{\label{fig: XPM_bd_sa}\includegraphics[width=0.15\textwidth]{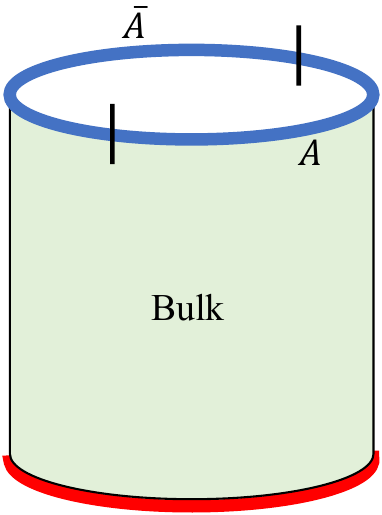}}
        \caption{\label{fig: RTPM_bd_trans} Boundary phase transition of RTPM.}
\end{figure*}

This phase transition is reflected in the scaling of $\mathrm{sym} I_{AB}$ and entropy $\mathrm{sym}S_A$ of the boundary symmetries associated with the free boundary. We find that the boundary symmetry $\mathrm{sym} I_{AB}$ of two antipodal boundary domains $A,B$ with $L_A = L_B = L/8$ collapses to a universal function
\begin{equation}
    \mathrm{sym}I_{AB}^{bd} \sim g((p_3 - p_3^c)L^{1/\nu_3'})
\end{equation}
with $p_3^c \sim 0.81$ and $\nu_3' \sim 2.43$, as shown in Fig.~\ref{fig: RTPM_bd_mi_dc}. This collapse indicates a boundary phase transition as we tune the coupling parameter $p_3$. Here the exponent $\nu_3'$ is related to $\nu_3 \sim 1.21$ obtained previously in the following manner
\begin{equation}
    \nu_3' = z \times \nu_3
\end{equation}
with $z = 1.697$ being the effective dynamical exponent extracted in the previous section.

When $p_3 > p_3^c$, we have $\mathrm{sym}I_{AB}^{bd} > 0$, indicating that the boundary is long-range correlated and the boundary symmetry operators are non-local. For $p_3 < p_3^{bd}$, $\mathrm{sym}I_{AB}^{bd} = 0$, meaning that the boundary symmetry operators are local and the boundary is short-range correlated. The boundary transition is further reflected in the scaling of half-boundary symmetry entropy
\begin{equation}
    \mathrm{sym}S_{L/2}^{bd}\sim \left\{
    \begin{matrix}
        L^a        &p_3 > p_3^c\\
        s_0        &p_3 = p_3^c\\
        \exp(-b L) &p_3 < p_3^c
    \end{matrix}
    \right.
\end{equation}
with $s_0 \sim 3$ and $a, b > 0$, as shown in Fig.~\ref{fig: RTPM_bd_ee_L}. For $p_3 > p_3^c$, the non-local symmetry operators contribute to the power-law scaling of the symmetry entropy, while at $p_3 < p_3^c$, the symmetry entropy is exponentially suppressed by system size $L$. In the active phase where $p_3 > p_3^{bd}$, $\mathrm{sym}S_{L/2}^{bd}$ exhibits a sublinear power-law scaling with $\alpha \sim 0.5$. This finding is consistent with the scaling behavior of the bulk symmetry operators on the cylinder geometry, indicating that while they are non-local, they are not extensive. This is different from the scaling behavior on the torus geometry as shown in Fig.~\ref{fig: op_size_XPM} and the physics of such a difference is left for the future work.

\begin{figure}[ht]
            \centering
            \includegraphics[width=0.3\textwidth]{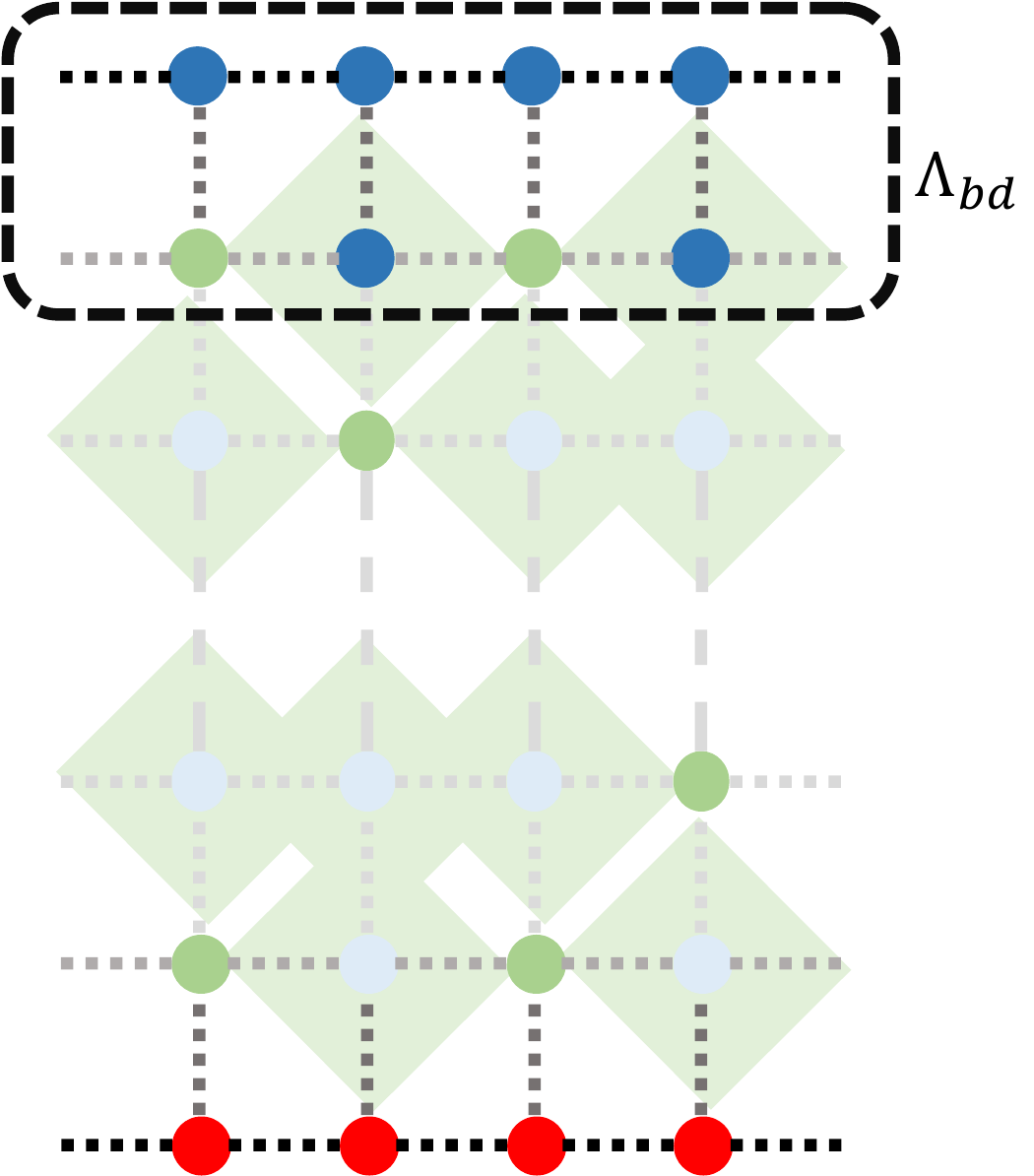}
            \caption{RXPM with free boundary on top (dark blue dots) and fixed boundary on bottom. The bulk five-body interactions are illustrated as green squares and the single-site terms are illustrated as green dots.}
            \label{fig: RXPM_bd}
        \end{figure}

    \paragraph{RPXM boundary transition}
        For RXPM, the free boundary is selected as shown in Fig.~\ref{fig: RXPM_bd}. Different from that of RTPM, it contains two layers of spins. Again, we fix the last layer of boundary spins to be $\sigma = +1$. 
    
        \begin{figure}[ht]
         \centering
         \subfloat[\label{fig: CAwRI0.8} $p_5 = 0.8$]{\includegraphics[width=0.12\textwidth]{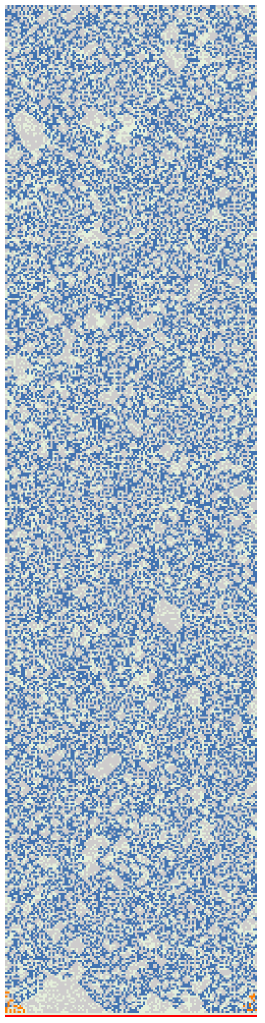}}\qquad
         \subfloat[\label{fig: CAwRI0.743} $p_5 = 0.743$]{\includegraphics[width=0.12\textwidth]{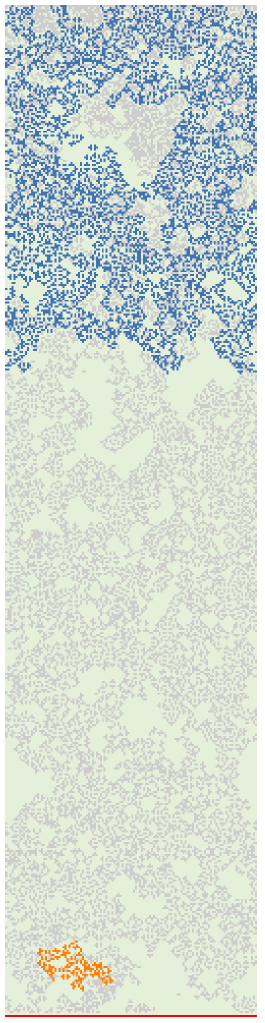}}\qquad
         \subfloat[\label{fig: CAwRI0.7} $p_5 = 0.7$]{\includegraphics[width=0.12\textwidth]{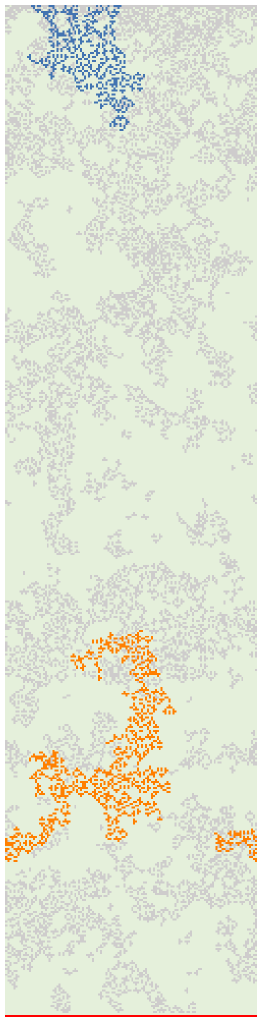}}\qquad
         \caption{\label{fig: RXPM_CAwRI_evo} Space-time configuration of the RXPM-CAwRI evolution. Time flows from top to bottom. Blue sites host spin-flip symmetries associated with the boundary, the orange sites carry spin-flips related to the bulk, and the grey sites support the full symmetry group.}
     \end{figure}

\begin{figure}[ht]
        \centering
        \subfloat[Geometry of half entropy $\mathrm{sym}S_{L/2}^{bd}$]{\label{fig: TPM_bd_sa}\includegraphics[width=0.2\textwidth]{figures/RTPM_bd_hf_ee_set.pdf}}\qquad
        \subfloat[$\mathrm{sym}S_{L/2}^{bd}$ of the RXPM]{\label{fig: TPM_bd_sa_R}\includegraphics[width=0.3\textwidth]{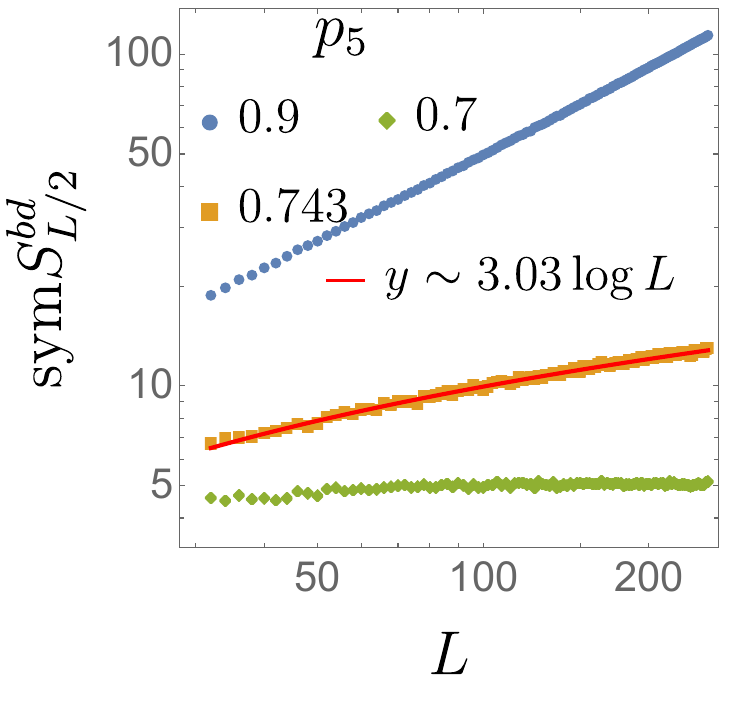}}
        \caption{Boundary symmetry entropy on the top boundary of RXPM.}
    \end{figure}

    We observed a boundary phase transition as we tune the bulk coupling parameter $p_5$, as present in Fig.~\ref{fig: TPM_bd_sa_R}. When the five-body bulk couplings dominate, $p_5 > p_5^c$, the symmetries in the system are extensive and non-local, resulting in a volume law scaling in the boundary symmetry entropy $\mathrm{sym}S_{L/2}^{bd} \sim L$. On the other hand, when $p_5 < p_5^c$, the symmetry operator breaks into small clusters with finite length, resulting in an area-law scaling of the boundary symmetry entropy $\mathrm{sym}S_{L/2}^{bd} \sim O(1)$.
    
\begin{figure}[ht]
        \centering
        \subfloat[Geometry of anti-podal boundary domain $A, B$ with $L_A = L_B = L / 8$]{\label{fig: XPM_bd_mi}\includegraphics[width=0.2\textwidth]{figures/bd_mi_diag.pdf}}\qquad
        \subfloat[Data collapse of $\mathrm{sym}I_{AB}^{bd}$ of RXPM]{\label{fig: XPM_bd_mi_dc}\includegraphics[width=0.3\textwidth]{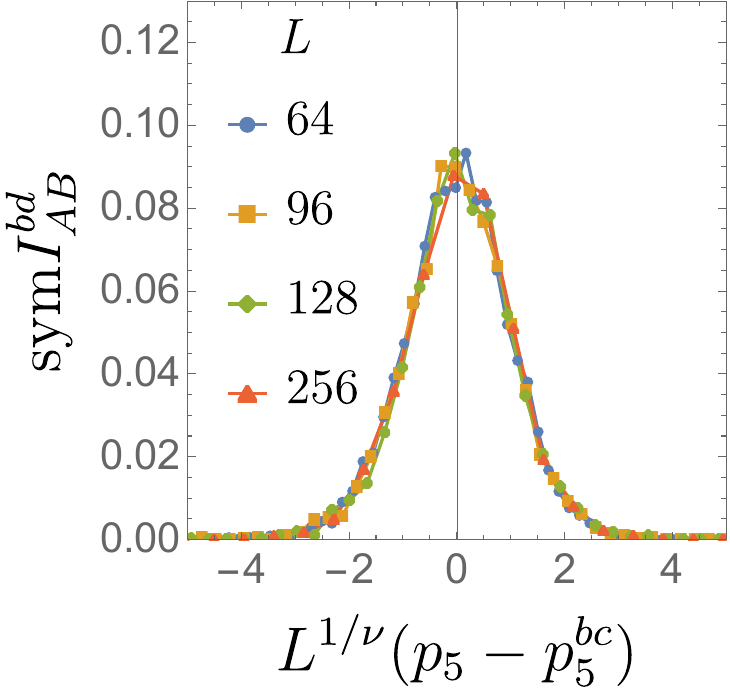}}\qquad
        \caption{RXPM $\mathrm{sym}I_{AB}^{bd}$ data collapse.}
    \end{figure}

    The critical point is further characterized by the peak of $\mathrm{sym}S_{L/2}^{bd}$ as we tune the coupling parameter $p_5$. As shown in Fig.~\ref{fig: XPM_bd_mi_dc}, $\mathrm{sym}S_{L/2}^{bd}$ collapses to a universal function
    \begin{equation}
        \mathrm{sym}I_{AB}^{bd} \sim h((p_5 - p_5^c)L^{1/\nu_5})
    \end{equation}
    with $p_5^c = 0.743$ and $\nu_5 = 1.3$.
    
\begin{figure}[ht]
        \centering
        \subfloat[Critical $\mathrm{sym}S_{A}^{bd}$ scaling of RXPM]{\label{fig: XPM_bd_ee}\includegraphics[width=0.3\textwidth]{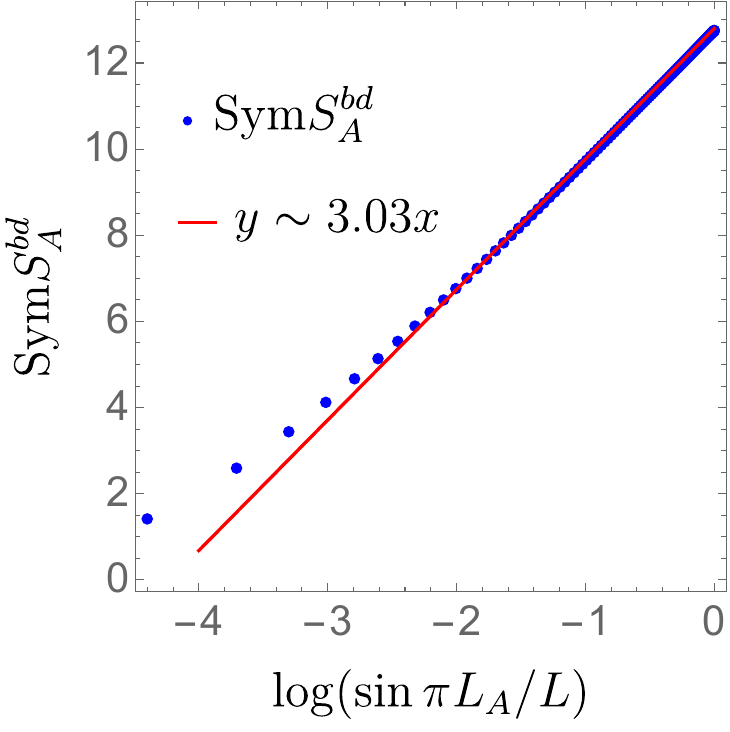}}\qquad
         \subfloat[Critical $\mathrm{sym}I_{AB}^{bd}$ scaling of RXPM]{\label{fig: RXPM_bd_mi_crit}\includegraphics[width=0.3\textwidth]{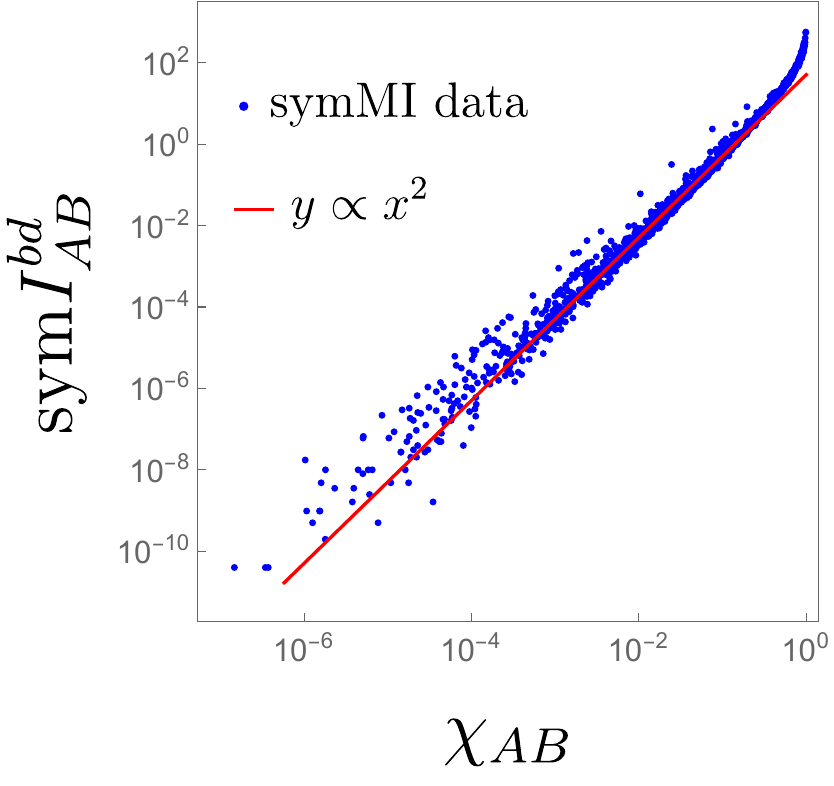}}\qquad
        \caption{RXPM boundary criticality data ($p_5 = 0.743$).}
    \end{figure}

    At criticality $p_5  = p_5^c = 0.743$, the symmetry entropy scales logarithmically
    \begin{equation}
        \mathrm{sym}S_A^{bd} \sim  c \log \sin \left(\frac{\pi L_A}{L}\right),
    \end{equation}
    with $c \sim 3.03$ as shown in Fig.~\ref{fig: XPM_bd_ee}. The symmetry mutual information $\mathrm{sym}I_{AB}^{bd}$ scales as
    \begin{equation}
        \mathrm{sym}I_{AB}^{bd} \sim \chi_{AB}^{\Delta}
    \end{equation}
    with $\Delta = 2$, and $\chi_{AB}$ being the cross-ratio 
    \begin{equation}\label{eq: cr}
        \chi_{AB} = \frac{x_{12}x_{34}}{x_{13}x_{24}},\quad \mbox{with}~x_{ij} = \frac{L}{\pi}\sin\left(\frac{\pi}{L}|x_i - x_j |\right),
    \end{equation}
    where $x_{i = 1,2,3,4}$ are endpoints of non-overlapping subregions $A = [x_1, x_2]$ and $B = [x_3, x_4]$ present in Fig.~\ref{fig: XPM_bd_mi}.  
    
\subsection{Plaquette models, stabilizer state measurement, and boundary entanglement structure}

    This part discusses the connection between RXPM, stabilizer state measurement, and the measurement-induced phase transition in hybrid Clifford dynamics. We show that Pauli measurements on a stabilizer state can be understood in terms of the Plaquette models, and its spin-flip symmetry characterizes the entanglement structure induced by the measurement. We use the cluster state defined on the square lattice as an example to illustrate this connection further. As previously shown in Refs.~\cite{PhysRevB.106.144311,liu2023quantum}, measuring the bulk qubits on the square lattice cluster state will induce an interesting entanglement structure on the boundary. In this section, we further show that the boundary spin-flip symmetry group of RXPM can fully characterize this entanglement structure.

    \subsubsection{Plaquette models and measuring the stabilizer state}


The stabilizer state is a special quantum state where all information is conveniently stored in a stabilizer tableau. It plays a critical role in simulating Clifford circuits, and an efficient classical algorithm for this is outlined in~\cite{ gottesman1998heisenberg}. 

A stabilizer state $\ket{\phi}_S$ adheres to the condition:
\begin{equation}
    \ket{\phi}_S = \mathbf{g} \ket{\phi}_S \quad \mathbf{g} \in \mathcal{S} \subset \mathcal{P}_N.
\end{equation}
Here, $\mathbf{g}$ is the group element of the stabilizer group $\mathcal{S}$, and $\mathcal P_N$ denotes the $N$-qubit Pauli group.
In the following, we represent such group by its generators $\{\mathbf{g}_{i=1,\ldots,N}\}$
\begin{equation}
    \mathcal{S} = \langle \mathbf{g}_1,\mathbf{g}_2,\ldots,\mathbf{g}_N \rangle 
\end{equation}
where $\langle \ldots \rangle$ denotes the group generated by `$\ldots$'.

Performing a Pauli measurement on such a quantum state induces a change in the stabilizer group $\mathcal{S} \to \mathcal{S}'$. To elucidate this change, we introduce the commutator matrix $P$:
\begin{equation}
    O_i \mathbf{g}_j = (-1)^{P_{ij}}\mathbf{g}_j O_i
\end{equation}
where $O_i = X_i, Y_i, Z_i$ is a Pauli support on site-$i$ serving as the measurement observable, and $g_j$ is the $j$-th generator of the pre-measurement stabilizer group $\mathcal{S}$. This $P$ matrix encapsulates the commutation relation between the pre-measurement stabilizer group and the measurement observable. The post-measurement stabilizer group $\mathcal{S}'$ comprises combinations of generators from the pre-measurement stabilizer group $\mathcal{S}$ and the measurement observables $\{O_i\}$. Formally:
\begin{equation}\label{eq: post_stab}
    S' = \big\langle O_i , ~\mathbf{g}'_k \equiv \prod_{j} \mathbf{g}_j^{x_{k}^{j}},~ \mathbf{g}_m\big\rangle.
\end{equation}
as the post-measurement quantum state collapse to one of the eigenstates of the observable $O_i$. Here $x_{k}^j = 0, 1$ characterizes the stabilizer generators induced by the measurement and $ \mathbf{g}_m$ denotes pre-measurement stabilizers that commute with all the observables, $[\mathbf{g}_m, O_i] = 0$ for all $i,m$. From here and in the remaining part of this section, we use $\prime$ to label the post-measurement stabilizer group and its elements. 

To maintain the abelian nature of the stabilizer group, it is imperative that
\begin{equation}
    [O_i, \mathbf{g}'_k] = 0,
\end{equation}
which can be further written as
\begin{equation}
    \big[1 - (-1)^{\sum_{j} P_{ij} x_{k}^{j}}\big]O_i\mathbf{g}'_k = 0,
\end{equation}
signifying that
\begin{equation}
    \sum_{j} P_{ij} x_{k}^{j} = 0 ~\operatorname{mod}~ 2 \quad \forall i, k.
\end{equation}
This equation is the same as Eq.~\ref{eq:const}, which is used to capture the ground state physics of the plaquette models. Taking advantage of this equation, we may directly write down the corresponding plaquette model Hamiltonian
\begin{equation}\label{eq: stab_plaq}
    H[\sigma] = -\sum_{q} J_q \prod_{i: P_{qi} = 1} \sigma_i,
\end{equation}
where the product $\prod_{i: P_{qi} = 1}$ is over non-trivial entries $i$ of the $q$-th row of the matrix $P$. The Pauli measurement process is thus related to the Plaquette models. Each of the post-measurement stabilizer
\begin{equation}\label{eq: stab_symm}
    \mathbf{g}'_k \equiv \prod_{j} \mathbf{g}_j^{x_{k}^{j}}
\end{equation}
is thus related to the spin-flip symmetry operator in the plaquette model via the vector $\vec{x}_k = (x_k^1, \ldots, x_k^N)$.

\subsubsection{Bulk measurement induced boundary entanglement}
\begin{figure}[ht]
    \centering
    \includegraphics[width=0.45\textwidth]{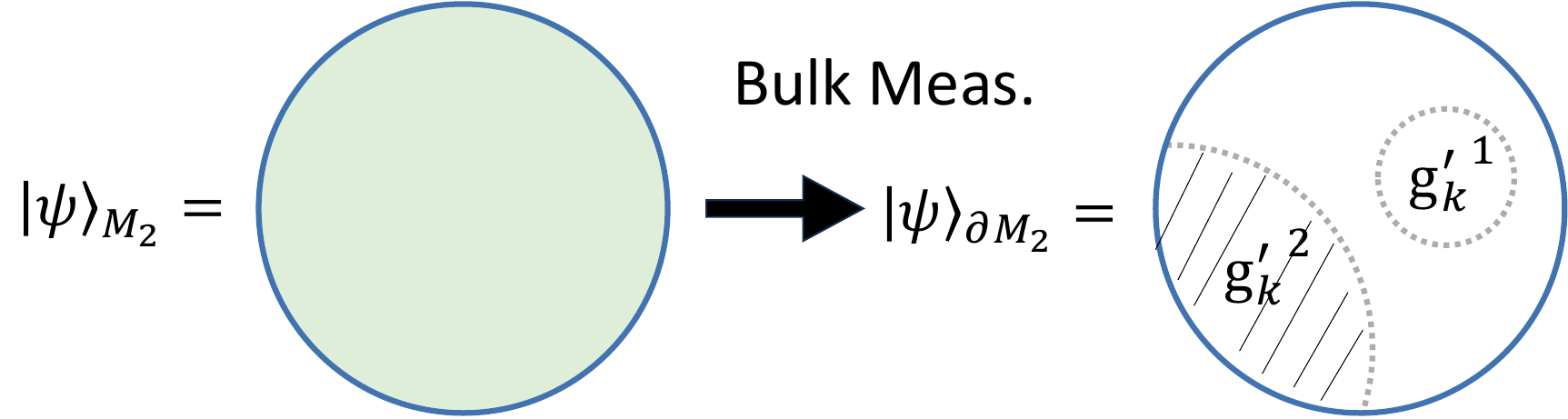}
    \caption{Measuring the bulk of a 2D quantum state on a 2D manifold results in a 1D quantum state living on the manifold boundary. The measurement-induced stabilizers separates into two types: 1. Bulk stabilizers ${\mathbf{g}_k'}^{1}$. 2. Boundary stabilizers ${\mathbf{g}_k'}^{2}$. The entanglement strcutre of the boundary state is captured by the boundary stabilizers ${\mathbf{g}_k'}^{2}$.}
    \label{fig: mbqc}
\end{figure}

    We now consider a specific setup where a $d$-dimensional quantum stabilizer state is obtained via measuring the bulk of a $d + 1$-dimensional stabilizer state.
    An example for $d = 1$ is presented in Fig.~\ref{fig: mbqc}. This type of system has been previously studied in~\cite{PhysRevB.106.144311, liu2023quantum, okuda2024anomaly, PhysRevB.109.125148}. These systems are also referred to as measurement-based quantum computers (MBQC)~\cite{PhysRevA.68.022312, PhysRevLett.86.5188}, as the entanglement structure of the post-measurement state is controlled by the bulk measurement direction. We now show that the entanglement structure of such a quantum system can be understood via the structure of the boundary spin-flip group of the corresponding plaquette models.

As shown in Eq.~\ref{eq: post_stab}, the post-measurement stabilizer group consists of three parts: the measurement observables $\{O_i\}$, the measurement-induced stabilizer $\mathbf{g}_k'$. In the MBQC setup, if we further assume that the bulk measurements are single-site Paulis $O_i = X_i, Y_i, Z_i$, and no stabilizer lives solely on the boundary, the corresponding stabilizer group is then
\begin{equation}
    \mathcal{S}' = \langle \mathbf{g}'_k, O_i \rangle.
\end{equation}
The non-trivial entanglement structure of the unmeasured boundary is thus captured by $\mathbf{g}'_k$'s. The $\{\mathbf{g}'_k\}$ are naturally separated into two types as present in Fig.~\ref{fig: mbqc}: 1) the bulk stabilizers, those support trivially on the boundary labeled by $\{{\mathbf{g}'_k}^1\}$ and 2) the boundary stabilizers, those have non-trivial support on the boundary labeled by $\{{\mathbf{g}'_k}^2\}$. Obviously, the first type can be directly decomposed into the bulk measurement observables 
\begin{equation}
    {\mathbf{g}'_k}^1 = \prod O_i.
\end{equation}
The post-measurement stabilizer group is then
\begin{equation}
    \mathcal{S}' = \langle {\mathbf{g}'_k}^2, O_i \rangle,
\end{equation}
and non-trivial boundary entanglement structure is then fully captured by the boundary part ${\mathbf{g}'_k}^2$.

Formally, the bulk stabilizers form a subgroup of the full stabilizer group ${\mathcal{S}^1}' = \{{\mathbf{g}'_k}^1\} \subset {\mathcal{S}}'$, and the boundary part ${\mathcal{S}^2}' = \{{\mathbf{g}'_k}^2\}$ is the group quotient
\begin{equation}
    {\mathcal{S}^2}' = \mathcal{S}'/{\mathcal{S}^1}'.
\end{equation}
Taking advantage of the relation between the post-measurement stabilizers and the spin-flip symmetry obtained in Eq.~\ref{eq: stab_symm}, and the bulk-boundary correspondence of symmetries given in Eq.~\ref{eq: g_bd}
$$G^{bd} = G/G^{bk},$$ 
it is now obvious that the bulk stabilizers $\mathcal{S}_1'$ correspond to the bulk spin-flip symmetry $G^{bk}$ and the boundary stabilizers $\mathcal{S}_2'$ then correspond to the boundary spin-flip symmetry $G^{bd}$.

Recall that the entanglement entropy of the stabilizer state~\cite{fattal2004entanglement}
\begin{equation}
     S_A = \frac{1}{2} \log |\mathcal{S}_{A \overline{A}}|
\end{equation}
where $\log |\mathcal{S}_{A \overline{A}}|$ counts the rank group quotient 
\begin{equation}
    \mathcal{S}_{A \overline{A}}\equiv\mathcal{S}/(\mathcal{S}_A\mathcal{S}_{\overline{A}})
\end{equation}
where $\mathcal{S}_{A/\overline{A}} \subset \mathcal{S}$ are the subgroup of stabilizers that operate only on domain $A/\overline{A}$.

\begin{figure}[ht]
     \centering
     \subfloat[2D domain: domain $A,A_0$ is enclosed by red/green line]{\label{fig:2domain}\includegraphics[width=0.2\textwidth]{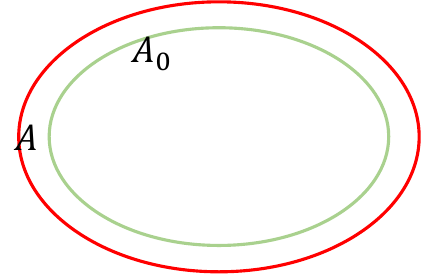}}\qquad
     \subfloat[1D boundary domain: $A,A_0$ is enclosed by the red/green dashed line]{\label{fig:1domain}\includegraphics[width=0.2\textwidth]{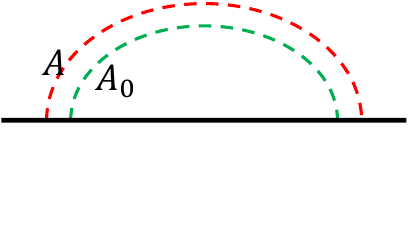}}
     \caption{\label{fig:domainA} Relation between domain $A$ and $A_0$.}
\end{figure}

As shown in Eq.~\ref{eq: stab_plaq}, the stabilizer group $\mathcal{S}'$ and the plaquette model spin-flip group $G$ are related via the binary vector set $\{x_k\}$. As by assumption $g_i$'s are local operators, the entanglement entropy of the boundary state $\ket{\psi}_{\partial M_2}$ is then related to the boundary symmetry entropy $\mathrm{sym}$ up to an area-law deviation in the following manner
\begin{equation}\label{eq: EE_symEE}
    S_A = \frac{1}{2} \text{sym} S_A^{bd} + O(l_A),
\end{equation}
where $l_A$ is the size of the boundary of domain $A$. For a 1D domain, its boundary $l_A \sim 1$. This deviation $O(l_A)$ originates from the difference in the local subgroup size $\log |G_A|$ and $\log |\mathcal{S}'_A|$, which can be directly calculated as
\begin{equation}
\begin{aligned}
     \delta_A &\equiv \log |G_A| - \log |\mathcal{S}'_A|\\
     &= \log |G_A| - \log |G_{A_0}|\\
     &= \rank T_{\overline{A_0}} - \rank T_{\overline{A}}\\
     &\le \rank T_{A\backslash A_0} \le \min\{|G|, l_A \} \\
     &\le l_A
\end{aligned}
\end{equation}
where in the second line $G_{A_0}$ is isomorphic to the local stabilizer group $\mathcal{S}_A'$ as generators $g_i$'s are local we have $A_0 \subset A$ and $|A \backslash A_0| \sim l_A$, and the geometry of domain $A$ and $A_0$ is presented in Fig.~\ref{fig:domainA}. In the third line, we used the fact that
\begin{equation}
    \log |G_{\mathcal{D}}| = \rank T - \rank T_{\overline{\mathcal D}}
\end{equation}
for any domain $\mathcal{D}$ with $N$ being the total number of sites. In the fourth line, we used the rank inequality that for any matrices $\mathbf{A}$ and $\mathbf{B}$ 
\begin{equation}
    \rank{\mathbf{A}} + \rank{\mathbf{B}} \ge \rank [\mathbf{A} \mid \mathbf{B}],
\end{equation}
and in the fifth line for any $a \times b$ matrix $\mathbf{A}$
\begin{equation}
    \rank \mathbf{A} \le \min\{a, b\}.
\end{equation}

\subsubsection{RXPM, MBQC boundary, and Clifford MIPT}

Building upon the mappings described above, we are now ready to show the equivalence between the boundary symmetry phase transition observed in RXPM and the measurement induced boundary phase transition on a two dimensional cluster state.

\begin{figure}
    \centering
    \includegraphics[width=0.25\textwidth]{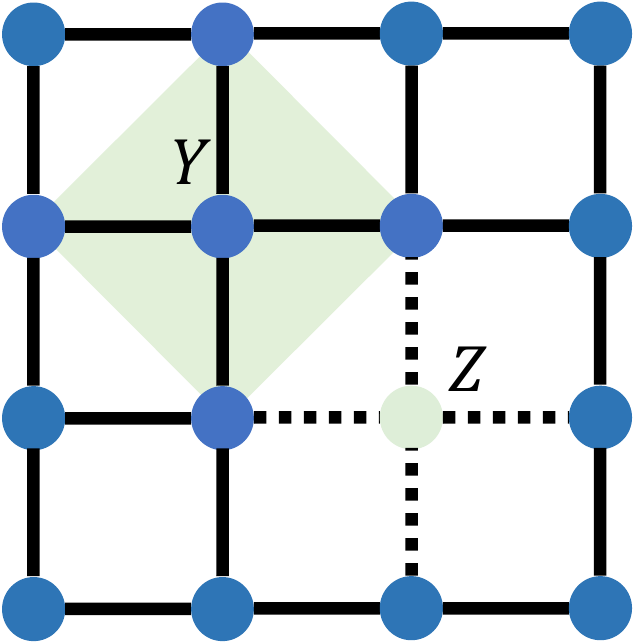}
    \caption{$Y$ measurement induces $5$ non-zero entries in the corresponding row of the matrix $P$, corresponding to site $i$ and its four neighboring sites (highlighted in Green). $Z$ measurement only induces one non-zero entry result in in the corresponding row of the matrix $P$ (green dot enclosed by the dashed line).}
    \label{fig: RXPM_mbqc}
\end{figure}

For the square lattice graph state, the stabilizer generators are given by
\begin{equation}
    g_i = X_i \prod_{j\in N_i} Z_j,
\end{equation}
where $N_i$ is the set of four neighboring sites of site $i$. We measure all of the bulk qubits in $Y$ or $Z$ directions. As present in Fig.~\ref{fig: RXPM_mbqc}, the single site $Y$ measurement anti-commutes with five $g_i$ and corresponds to the five-body interaction. On the other hand, the $Z$ measurement corresponds to the single-site term, as this single $Z$ only anti-commutes with one $g_i$.

The generators of the post-measurement stabilizer state corresponds to the boundary symmetry operators of RXPM. In addition, the entanglement entropy of the boundary stabilizer state is equivalent to half of the boundary symmetry entropy $\mathrm{sym}S^{bd}_A$ with respect to subregion $A$ up to a finite constant.

The critical exponents extracted in the RXPM model (this work), the Y/Z bulk measurement-induced boundary criticality on a 2D square lattice cluster state~\cite{liu2023quantum}, and the measurement-induced phase transition in random Clifford circuits~\cite{li2019measurement, skinner2019measurement, gullans2020dynamical} are shown in Tab.~\ref{tab: exp}. We propose that these criticalities belong to the same universality class, supported by the proximity of exponents. Here, $c$ and $h^0$ for RXPM and Clifford MIPT/MBQC (square) differ by a prefactor $2$. The reason lies in the definition of stabilizer entanglement entropy and symmetry entropy shown in Eq.~\ref{eq: EE_symEE}.

\begin{table}[ht]
    \begin{tabular}{|l|*{6}{p{0.8cm}|}}
    \hline
    Model       & $z$ & $h^0$ & $h^1$   & $\nu$  & $c$    & $\Delta$ \\ \hline
    RXPM        & $1$ & $1.53$& $0.125$ & $1.3$  & $3.05$ & $2$      \\ \hline
    MBQC(Square)& $1$ & -     & -       &  $1.3$ & $1.6$  & $2$      \\ \hline
    Cliff. MIPT & $1$ & $0.76$ & $0.125$ & $1.3$ & $1.6$  & $2$      \\ \hline
    \end{tabular}
    \caption{\label{tab: exp} Critical exponents of RXPM boundary, random Y/Z measurement MBQC on the square-lattice boundary \cite{liu2023quantum} and Clifford MIPT \cite{li2019measurement, skinner2019measurement, gullans2020dynamical}. Here, we drop the subscript of $\nu_5$ for the RXPM model.}
\end{table}

\section{Conclusion and discussion}\label{sec: conclusion}

In this work, we introduce a class of classical spin models with $q$-spin Ising interactions defined on the two dimensional lattice. These interactions are randomly replaced with single-site terms with probability $1-p$, leading to what we term random plaquette models (RPMs). We investigate the ground state phase transitions of these models by analyzing the behavior of the spin-flip symmetry operators. Specifically, we concentrate on two instances: the random triangular plaquette model (RTPM) with $q=3$ and the random X-plaquette model (RXPM) with $q=5$. Our analysis reveals that for $p>p_c$, the symmetry operators become non-local, spanning the entire lattice. Conversely, for $p<p_c$, the symmetry operator localizes within a finite region. To characterize these localization phase transitions, we develop various tools tailored to the behavior of these symmetry operators.

We also present a dynamical perspective to elucidate the phase transitions in the localization of the symmetry operator. We demonstrate that the symmetry operator, when initiated from a one-dimensional boundary, evolves according to an update rule determined by the specific interaction form. By employing this approach, we construct classical 1+1D random cellular automata and illustrate that the propagation of the symmetry operators, starting from one boundary and moving into the bulk, can undergo an absorbing phase transition. In particular, we establish that the RTPM corresponds to a special limit of the Domany-Kinzel model. Regarding the RXPM, we determine that it possesses a dynamical exponent $z=1$ at criticality and exhibits an intriguing connection with the recently discovered measurement-induced entanglement phase transition (MIPT).

The connection between them is twofold. First, our model offers a classical counterpart to the MIPT, akin to other proposals discussed in \cite{Iaconis_2020,Han_2023,Lyons_2023,Willsher_2022, kelly2023coherence}. Here, the single-site constraint plays the role of the measurement in our model. Second, we establish an equivalence between the boundary symmetry phase transition of the RXPM and the boundary entanglement phase transitions induced by the bulk measurement in the cluster state defined on the square lattice. The latter can be effectively treated as the entanglement phase transition in a 1+1D hybrid Clifford circuit \cite{li2019measurement}, as previously demonstrated in \cite{liu2023quantum}. By numerically comparing the critical exponents in these three transitions, we demonstrate that the phase transitions in these three models belong to the same universality class, as presented in Tab.~\ref{tab: exp}.

We conclude this work by highlighting several promising directions for further research. Firstly, the physics explored in this paper can be generalized to other 2D models with multi-spin interactions or even some non-local spin models. It would be interesting to explore the universality classes of potential phase transitions in these spin models. 

Secondly, the symmetry operator defined in our classical spin model is equivalent to the logical operator in classical LDPC codes~\cite{pless1998introduction,rakovszky2023physics}. Exploring our transition from the perspective of classical error correction and even promoting it to the quantum version could be an interesting direction. 

Thirdly, our model offers a new theoretical framework for understanding the MIPT in hybrid Clifford circuits. Generalizing our approach to more complex classical spin models could deepen our understanding of the MIPT in Haar random circuits, which is currently only comprehended in specific limits or at a mean-field level~\cite{skinner2019measurement,PhysRevB.101.104302,Bao_2020}. 

Fourthly, it is known that the Newman-Moore model with three-body interaction or, more generically, the plaquette models do not have a thermal phase transition but do exhibit interesting glassy dynamics at finite temperature~\cite{PhysRevE.60.5068, PhysRevE.62.7670,jack2005caging,JuanPGarrahan_2002,yamaguchi2010static,physchem_040808_090405}. Substitute the multi-body interaction terms with single-site terms would suppress the glassy behavior at finite temperatures. Therefore, it would be interesting to explore the finite temperature physics of the models discussed in this paper and their potential glass phase transitions.

Lastly, our model bears some resemblance to the XOR-SAT problem, where two transitions have been identified by varying the number of random non-local constraints~\cite{doi:10.1126/science.264.5163.1297, mezard2003two}. Our model appears to be a local version of XOR-SAT, and exploring the connections between these two models could be interesting.

\begin{acknowledgments}

    We gratefully acknowledge computing resources from Research Services at Boston College and the assistance provided by Wei Qiu. H.L. thanks Ethan Lake, Shengqi Sang, Yabo Li, Konstantinos Sfairopoulos and Tianci Zhou for insightful discussions. This research is supported in part by the National Science Foundation under Grant No. DMR-2219735.
    
\end{acknowledgments}

\bibliography{ref}


\end{document}